\UseRawInputEncoding
\documentclass[aps,pre,reprint,superscriptaddress,letterpaper,twocolumn]{revtex4-1}
\usepackage{amssymb}
\usepackage{amsbsy}
\usepackage{amsmath}

\usepackage{bm}

\usepackage{graphicx}

\usepackage{grffile}
\usepackage{subfigure}

\usepackage{hyperref}
\usepackage{geometry}
\usepackage{bm}
\usepackage{lipsum} 
\usepackage{xcolor}

\usepackage{tikz}

\usepackage{eqnarray}

\begin{document}

\preprint{APS/123-QED}

\title{Global analysis of more than 50,000 SARS-Cov-2 genomes reveals epistasis between 8 viral genes}

\author{Hong-Li Zeng}
\email{hlzeng@njupt.edu.cn}
\affiliation{School of Science, and New Energy Technology Engineering Laboratory of Jiangsu Province, Nanjing University of Posts and Telecommunications, Nanjing 210023, China}
\affiliation{Nordita, Royal Institute of Technology, and Stockholm University, SE-10691 Stockholm, Sweden}

\author{Vito Dichio}%
\affiliation{Department of Physics, University of Trieste, Strada Costiera 11 34151, Trieste, Italy}
\affiliation{Nordita, Royal Institute of Technology, and Stockholm University, SE-10691 Stockholm, Sweden}
\affiliation{KTH -- Royal Institute of Technology, AlbaNova University Center, SE-106 91 Stockholm, Sweden}%

\author{Edwin Rodr{\'i}guez Horta}
\affiliation{Group of Complex Systems and Statistical Physics. Department of
Theoretical Physics, Physics Faculty, University of Havana, Cuba}

\author{Kaisa Thorell}
\affiliation{Department of Infectious Diseases, Institute of Biomedicine, Sahlgrenska Academy of the University of Gothenburg, Gothenburg, Sweden}
\affiliation{Center for Translational Microbiome Research, Department of Microbiology, Cell and Tumor biology, Karolinska Institutet, Stockholm, Sweden}

\author{Erik Aurell}
\email{eaurell@kth.se}
\affiliation{KTH -- Royal Institute of Technology, AlbaNova University Center, SE-106 91 Stockholm, Sweden}%

\date{\today}

\begin{abstract}

Genome-wide epistasis analysis is a powerful tool to infer gene interactions, which can guide drug and vaccine development and lead to deeper understanding of microbial pathogenesis. We have considered all complete SARS-CoV-2 genomes deposited in the
GISAID repository until \textbf{four} different cut-off dates, and used
Direct Coupling Analysis together with an assumption of
Quasi-Linkage Equilibrium to infer epistatic
contributions to fitness from polymorphic loci.
We find \textbf{eight} interactions, of which
three between
pairs
where one locus lies in gene ORF3a,
both loci holding non-synonymous mutations.
We also find interactions
between two loci in gene nsp13, both holding non-synonymous mutations,
and four interactions involving one locus holding a synonymous
mutation. Altogether we infer interactions
between loci in viral genes
ORF3a and nsp2, nsp12 and nsp6,
between ORF8 and nsp4,
and between loci in genes nsp2, nsp13 and nsp14.
The paper opens the prospect to use prominent
epistatically linked pairs
as a starting point to search for
combinatorial weaknesses of recombinant viral pathogens.
\end{abstract}

\keywords{SARS-CoV-2 $|$ Epistasis $|$ Recombination $|$ Direct Coupling Analysis} 
\maketitle


\section{Introduction}\label{sec:Introduction}

The pandemic of the disease COVID-19 has so far led to the confirmed deaths of more than 991,224 people~\cite{WHO} and has hurt millions.
As the health crisis has been met by Non-Pharmacological Interventions~\cite{Kraemer2020,Salje2020} there has been significant economic disruption in many countries. The search for vaccine or treatment against the new coronavirus SARS-CoV-2 is therefore a world-wide priority. The GISAID repository~\cite{GISAID} contains a rapidly increasing collection of SARS-CoV-2 whole-genome sequences, and has already been leveraged to identify mutational hotspots and potential drug targets~\cite{Pacchetti2020}. Coronaviruses in general exhibit a large amount of recombination~\cite{LaiCavanagh1997,Graham2010,Gribble2020,Li2020}. The distribution of genotypes in a viral population can therefore be expected to be in the state of Quasi-Linkage Equilibrium~\cite{Kimura1965,NeherShraiman2009,NeherShraiman2011}, and directly related to epistatic contributions to fitness~\cite{Gao2019,Zeng2020}. We have determined a list of the largest such contributions from 51,676 SARS-CoV-2 genomes by a Direct Coupling Analysis (DCA)~\cite{MorcosE1293,Cocco2018}.
This family of techniques has earlier been used to
infer the fitness landscape of HIV-1 Gag~\cite{Shekhar2013,Mann2014}
to connect bacterial genotypes and phenotypes through
co-evolutionary landscapes~\cite{Cheng2016}
and to enhance models of amino acid sequence evolution~\cite{delaPaz2020}.
We apply a recent enhancement of this technique to
eliminate
predictions that can be attributed to phylogenetics (shared inheritance)~\cite{Horta2019}. We find that eight predictions stand out between pairs of polymorphic sites located in genes nsp2 and ORF3a, nsp4 and ORF8, and between genes
nsp2, nsp6, nsp12, nsp13, nsp14 and ORF3a. Most of these sites have been documented in the literature when it comes to
single-locus variations~\cite{Forster2020,Khailany2020,Cai2020,Deng2020,Phelan2020,Sashittal2020}.
The nsp4-ORF8 pair
was additionally found to be strongly correlated
in an early study~\cite{Tang2020}. It does not
show prominent correlations today, but is ranked
second in our global analysis.
The epistasis analysis of this paper brings a different
perspective than correlations, and highlights pairwise
associations that have remained stable as orders of more SARS-CoV-2 genomes have been sequenced.

\section{Data and Methods}\label{sec:Data_Methods}
\subsection{Genome Data of SARS-CoV-2}
We analyzed the consensus sequences deposited in the GISAID database~\cite{GISAID} with high quality and full lengths (number of bps $\approx 30,000 $). Four data-sets are used for our investigation according to the collection date in GISAID database. The dates are 2020-04-01, 2020-04-08, 2020-05-02
and 2020-08-08 respectively.
The list of GISAID sequences used is
available on the Github repository~\cite{Zeng-github}.
The numbers of selected genomes are 2,268, 3,490, 10,587 and 51,676 
for each collection date.

\subsection{Multiple-Sequence Alignment (MSA)}
Multiple sequence alignments were constructed with the online alignment server MAFFT~\cite{Katoh2017,Kuraku2013} for the two smaller data sets with cut-off dates 2020-04-01 and 2020-04-08. To align the two larger data-set with more than 10,000 sequences, a pre-aligned reference MSA is recommended to accelerate the alignment and reduce the burden on computational resources. Here, we took the collection with cut-off date 2020-04-08 as the pre-aligned reference MSA for the two largest data set with cut-off dates 2020-05-02 and 2020-08-08.
The MSAs used are
available on the Github repository~\cite{Zeng-github}.

The MSA is a big matrix $\mathbf{S}=\{\sigma_i^n|i=1,...,L, n = 1,...,N\}$, composed of $N$ genomic sequences which are aligned over $L$ positions \cite{Cocco2018, Horta2019}. Each entry $\sigma_i^n$ of matrix $\mathbf{S}$ is either one of the 4 nucleotides (A,C,G,T), or ``not known nucleotide'' (N), or the alignment gap `-' introduced to treat nucleotide deletions or insertions, or some minorities.

\subsection{MSA filtering} Before filtering, we transform the MSA in two different ways as follows:
\begin{itemize}
    \item The gaps `-' are transformed to `N' while the minors `KFY...' are mapped to `N'. There thus 5 states remains, where `NACGT' are represented by `12345';
    \item The minors `KFY...' are mapped to `N'. Then there are 6 states, with `-NACGT' represented by `012345'.
\end{itemize}

The following criteria are used for pre-filtering of the MSA from the 2020-08-08 data-set. If for one locus the same nucleotide is found in more than 96.5\% of this column, or if the sum of the percentages of A, C, G and T at this position is less than 20\%, then this locus will be deleted. For each sequence, if the percentage of a certain nucleotide is more than 80\%, or if the sum of the percentages of A, C, G and T in this sequence is less than 20\%, then this sequence will be deleted.
With this filtering criteria, many loci but no sequences are deleted.
There are left 51,676 sequences and 689 loci.

\subsection{B-effective number}
\begin{figure}[htpb]
    \centering
    \includegraphics[width=0.4\textwidth]{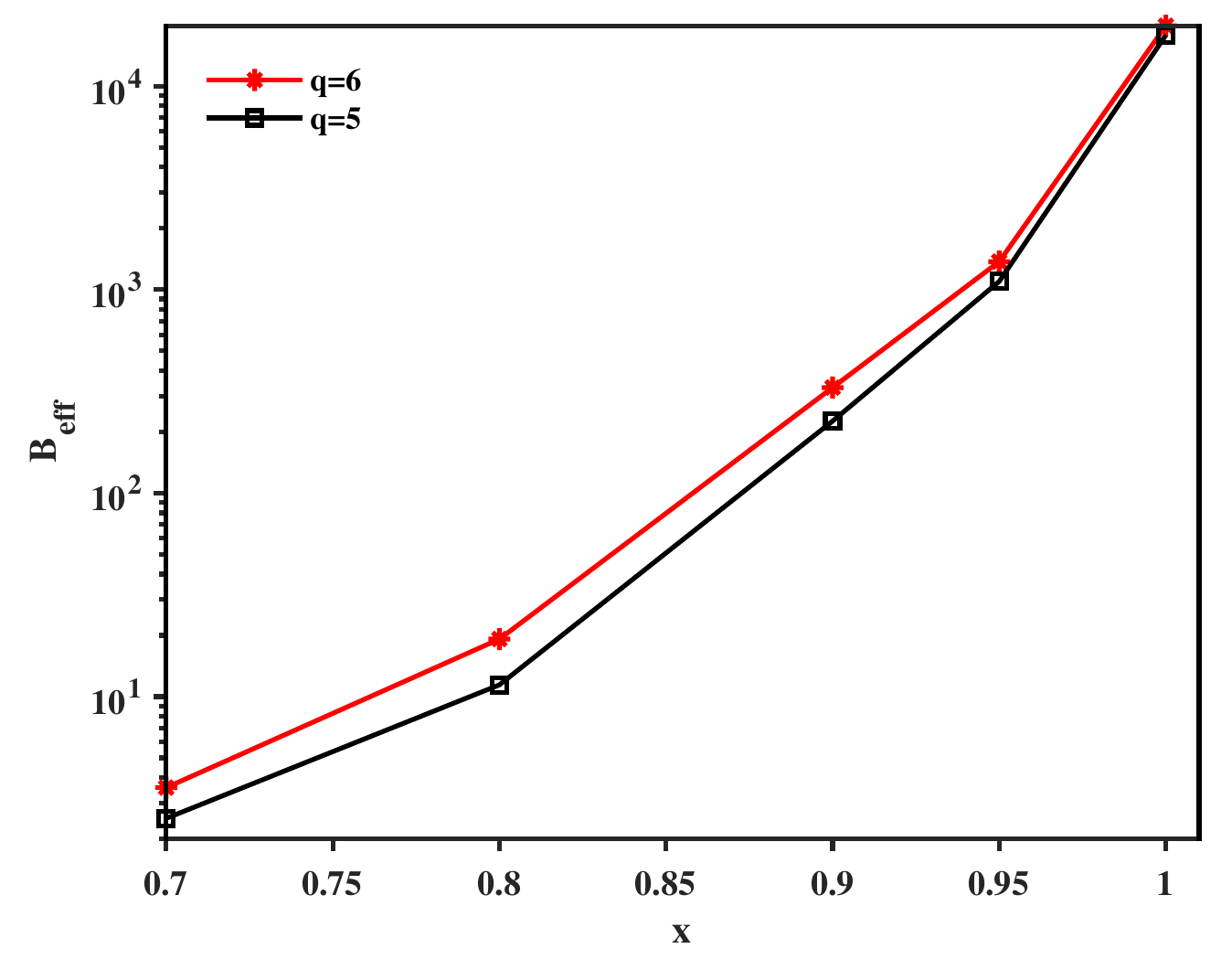}
    \caption{$B_{eff}$ number of the 2020-08-08 pre-filtered data-set with threshold $x$. Red for $q=6$ states (`-NACGT') and black for $q=5$ states (`NACGT'). The number of states are determined by the transform criteria of the pre-filtered MSA.}
    \label{fig:B-eff-num-x}
\end{figure}

To mitigate the effects of dependent samplings, it is standard practice to attach to each collected genome sequence $\sigma^{(b)}$ a weight $w_b$ \cite{MorcosE1293,Ekeberg-2014a,Cocco2018}, which normalizes its impact on the inference procedure.
An efficient way to measure the similarity between two sequences $\sigma^{(a)}$ and $\sigma^{(b)}$ is to compute the fraction of identical nucleotides and compare it with a preassigned threshold value $x$ in the range $0 \le x \le 1$.
The weight of a sequence $\sigma^{(b)}$ can be set as $w_b=\frac{1}{m_b}$, with $m_b$ the number of sequences in the MSA that are similar to $\boldsymbol{\sigma}^{(b)}$:
\begin{equation}
 m_b = |\{a\in\{1,...,B\}\}\texttt{:overlap}(\boldsymbol{\sigma}^{(a)},\boldsymbol{\sigma}^{(b)})\ge x|;
 \label{eq:dist}
\end{equation}

here \texttt{overlap} is the fraction of loci where the two sequences are identical.
The B-effective number of the transformed sequences is defined as
\begin{equation}
 B_{eff} = \sum_{b=1}^B w_b.
 \label{eq:B-eff}
\end{equation}
We compare the $B_{eff}$ value with different $x$ for the filtered MSA with $q=5$ and $q=6$ respectively. As shown in Fig.~\ref{fig:B-eff-num-x}, the data-set with 6 states shows larger $B_{eff}$ number for all tested $x$. We thus perform our analysis on the data-set with $q=6$ states, where `-NACGT' represented by `012345'.

The re-weighting procedure partially addresses a point raised
~\cite{MacLean2020}, that sequenced viral genomes are not a random sample of the global population. That is, even if sequencing is biased by the country they occur in and by contact tracing, sufficiently
similar genomes will have lower weight and so each will contribute less
to predictions.

\subsection{Elements of Quasi-Linkage-Equilibrium (QLE)} The
phenomenon of QLE was discovered by M.~Kimura
while investigating the steady-state
distribution over two bi-allelic loci evolving under
mutation, recombination and selection, with both
additive and epistatic contributions to fitness~\cite{Kimura1965}.
In the absence of epistasis such a system
evolves toward Linkage Equilibrium (LE) where the
distribution of alleles at the two loci are
independent. The covariance of alleles at the
two loci then vanishes. In the presence of
pairwise epistasis and sufficiently high rate of
recombination, the steady-state distribution
takes form of a Gibbs-Boltzmann form
\begin{equation}
 P(\sigma_1,...,\sigma_L)=\frac{1}{Z}\exp\{-H(\sigma_1,...,\sigma_L)\},
 \label{BoltzmannEQ}
\end{equation}
with an "energy function"
\begin{equation}
 H(\sigma_1,...,\sigma_L) = \sum_{i}h_i(\sigma_i) + \sum_{ij} J_{ij}(\sigma_i,\sigma_j)
 \label{Ising-Potts}
\end{equation}
In above $J_{ij}$ can be related to the
epistatic contribution to fitness between loci
$i$ and $j$ with alleles $\sigma_i$ and $\sigma_j$~\cite{NeherShraiman2009,NeherShraiman2011,Gao2019}.
The quantity $h_i$ is similarly a function of allele $\sigma_i$
which depends on both additive and epistatic contributions
to fitness involving locus $i$.
It has been verified in \textit{in silico} testing
that when the terms in \eqref{Ising-Potts} can be recovered
this is a means to infer epistatic fitness from
samples of genotypes in a population~\cite{Zeng2020}.
In the bacterial realm this approach was used earlier
to infer epistatic contributions to fitness in
the human pathogens \textit{Streptococcus pneumonia}~\cite{Skwark-2017a}
and \textit{Neisseria gonorrhoeae}~\cite{Schubert2019},
both of which are characterized by a high rate of
recombination.
The method was also tested on data on the bacterial pathogen \textit{Vibrio parahemolyticus}~\cite{Cui2019}.
In that study
the results from DCA were not superior from an analysis based on
Fisher exact test, see Appendix \ref{app:quantification_correlation} for a discussion.
This is consistent with the approach taken here, as
 \textit{V. parahemolyticus} has low rate of recombination.
Further details on the QLE state of evolving populations are
given in Appendix \ref{app:QLE}.

\subsection{Inference Method for epistasis between loci}

The basic assumption of modeling the filtered MSA is that it is composed by independent samples that follows the Gibbs-Boltzmann distribution~\eqref{BoltzmannEQ}
with $H$ as in \eqref{Ising-Potts}
Higher order interactions are also possible to include, but we ignore them here \cite{Schneidman2006}.
This assumption is a simplification of the biological reality, however provides an efficient way to extract information from massive data.

On the other hand, in the context of inference from protein sequences, it has been argued that the one encoded in Eq. (\ref{BoltzmannEQ}-\ref{Ising-Potts}) is the minimal \emph{generative} model i.e. capable not only to reproduce the empirical frequencies and correlations but also to generate new sequences indistinguishable from natural sequences \cite{Cocco2018,russ2005natural,socolich2005evolutionary}.

Many techniques have been developed to infer the direct couplings in Eq. (\ref{BoltzmannEQ}), as reviewed in \cite{Nguyen2017} and references therein,
see also Appendix \ref{app:DCA_methods}.
We employ the maximum pseudo-likelihood (PLM) method \cite{Besag-1975a,Ravikumar-2010a,Aurell-2012a,Ekeberg-2013a, Ekeberg-2014a, Gao2019} to infer the epistatic effects between loci from the aligned MSA.  PLM estimates parameters from conditional probabilities of one sequence conditioned on all the others. For Potts model with multiple states $q>2$, this conditional probability is

\begin{equation}
 P(\sigma_i|\boldsymbol{\sigma}_{\backslash i})=\frac{\exp\left(h_i(\sigma_i) +\sum_{j\ne i}J_{ij}(\sigma_i, \sigma_j)\right)}{\sum_{\mathbf{u}}\exp\left(h_i(u) +\sum_{j\ne i}J_{ij}(u, \sigma_j)\right)},
 \label{conditional_prob}
\end{equation}
with $\mathbf{u}=\{0,1,2,3,4,5\}$ the possible state of $\sigma_i$. Eq. (\ref{conditional_prob}) depends on a much smaller parameter set compared with that in Eq. (\ref{BoltzmannEQ}). This leads a much faster inference procedure of parameters compared with the maximum likelihood method. With a given independent sample sets, one can maximize the corresponding log-likelihood function
\begin{widetext}
\begin{equation}
       PL_i\left(h_i,\{J_{ij}\}\right)=  \frac{1}{N}\sum_s h_i\left(\sigma_i^{(s)}\right) +  \frac{1}{N}\sum_s \sum_{j \ne i} J_{ij}(\sigma_i^{(s)},\sigma_j^{(s)}) -\frac{1}{N}\sum_s \log\sum_{\mathbf{u}}\exp\big(h_i(u) +\sum_{j\ne i}J_{ij}(u,\sigma_j^{(s)})\big),
\end{equation}
\end{widetext}
where $s$ labels the sequences (samples), from $1$ to $N$.
With the filtered MSA, we then run the asymmetric version of PLM \cite{Ekeberg-2014a} in the implementation PLM available on~\cite{Gao-github} with regularization parameter $\lambda=0.1$. The inferred interactions between loci $i$ and $j$ are scored by the Frobenius norm.

\subsection{Relation to correlation analysis} In LE
the distributions of alleles over different loci
are independent. Given unlimited data and
unlimited computational resources, the terms $J_{ij}$
in \eqref{Ising-Potts} inferred from the data would then be zero.
The locus-locus co-variances,
defined as
\begin{equation}
 c_{ij}(a,b)=\left<\mathbf{1}_{\sigma_i,a}\mathbf{1}_{\sigma_j,b}\right>-\left<\mathbf{1}_{
\sigma_i,a}\right>\left<\mathbf{1}_{\sigma_j,b}\right>
 \label{LD}
\end{equation}
would also be zero.
The Frobenius norm of $c_{ij}(a,b)$ over
indices $(a,b)$ as a score of strength of correlations
would be zero as well.
The qualitative difference between correlation analysis
and global model inference based on
\eqref{BoltzmannEQ} and \eqref{Ising-Potts}
is that two loci $i$ and $j$ may be correlated ("indirectly coupled")
even if their interaction $J_{ij}$ is zero, provided they
both interact with a third locus $k$.
Data in Table~\ref{tab:tab4}
and Fig.~\ref{fig:corr-PLM-score-decrease}
show that the leading interactions retrieved by
DCA cannot be stably recovered in correlation analysis.
A different score of statistical dependency between two
categorical random variables is mutual information (MI).
Appendix \ref{app:quantification_correlation} shows that the
result does not substantially change if using
MI instead of Frobenuis norm of correlation matrices.
Circos plots of interactions based on correlation scores are
available on \cite{Zeng-github}.

\subsection{Epistasis analysis with PLM scores}  PLM procedure yields a fully connected paradigm between pairwise loci. To extract important information form massive SARS-CoV-2 genomic sequences, we focus on the significant scores between loci, the top-200 pairs. With a reference sequence ``Wuhan-Hu-1'', we identify the positions of the corresponding nucleotides. The visualization of these epistasis is performed by `circos' software~\cite{Krzywinski-2009a}.

\subsection{Randomized background distributions}
A way to assess the validity
of a small number of leading retained predictions
among a much larger set of mostly discarded predictions
is to compare to randomized backgrounds.
The retained predictions are then in any case large
(by some measure) and would also be retained if
selection would be made according to some cut-off,
or an empirical $p$-value.
The problem is thus how to distinguish
the case where a small sub-set of retained
values are large because they are different,
from the case when in a large number of samples
such values would appear at random.
This problem can be addressed by comparing the retained values
to the largest values from the same procedure
applied to randomized data, as was done
for predicted RNA-RNA binding energies
in a non-coding RNA discovery pipeline~\cite{Mandin2007}.
In the context of DCA (PLM) applied to genome-scale
MSAs, two earlier studies relying on randomized
background distributions are
\cite{Xu2018} and \cite{Gao2019}.

\subsection{PLM scores with randomization}
To understand  the nature of the top-200 PLM scores we perform two distinct randomization strategies on the MSA, such that its conservation patterns and (or) phylogenetic relations are preserved, while intrinsic co-evolutionary couplings (epistatic interactions) are removed~\cite{RodriguezWeigt-2020}.  Running DCA on artificial sequences ensembles generated by these strategies, and comparing them to the results obtained from original MSA allows to disentangle spurious couplings given by finite-size effects or by  phylogeny. The first strategy, we refer as 'profile',  randomizes the input MSA by random but independent permutation of all its columns conserving the single-columns statistics for all sites. This destroys all kind of correlations and  DCA couplings inferred from such samples are only non-zero due to the noise caused by finite sample-size. In the second strategy referred as 'phylogeny', the original MSA is randomized by a simulated annealing schedule where columns and rows are changed simultaneously  but so that inter-sequence distances are kept invariant. Phylogeny inferred from inter-sequence distance information would therefore be unchanged. Conversely, if the predicted epistatic interactions are due to phylogeny, they should also show up in terms recovered by PLM from MSAs scrambled by 'phylogeny'.
More details on the randomization strategies can be found in Appendix \ref{app:randomization}.

\section{Results}\label{sec:Results}
The predicted effective interactions between loci were obtained from Pseudo-Likelihood Maximization (PLM) scores, a standard computational
method to perform DCA. Manual inspection shows that about half of the top-50 links
and most of the top-200
involve noncoding sites in the 5' or 3' region on the ``Wuhan-Hu-1'' \cite{Wuhan-Hu-1} reference sequence, many of them have very short range and most of them with a large fraction of the gap or N (unknown nucleotide) symbols (data available
on~\cite{Zeng-github} for other data-set). We present the links with both terminal loci located in coding regions and the mutations  excluding gaps or `N's.

\begin{figure}[htpb]
\centering
\includegraphics[width=1.\linewidth]{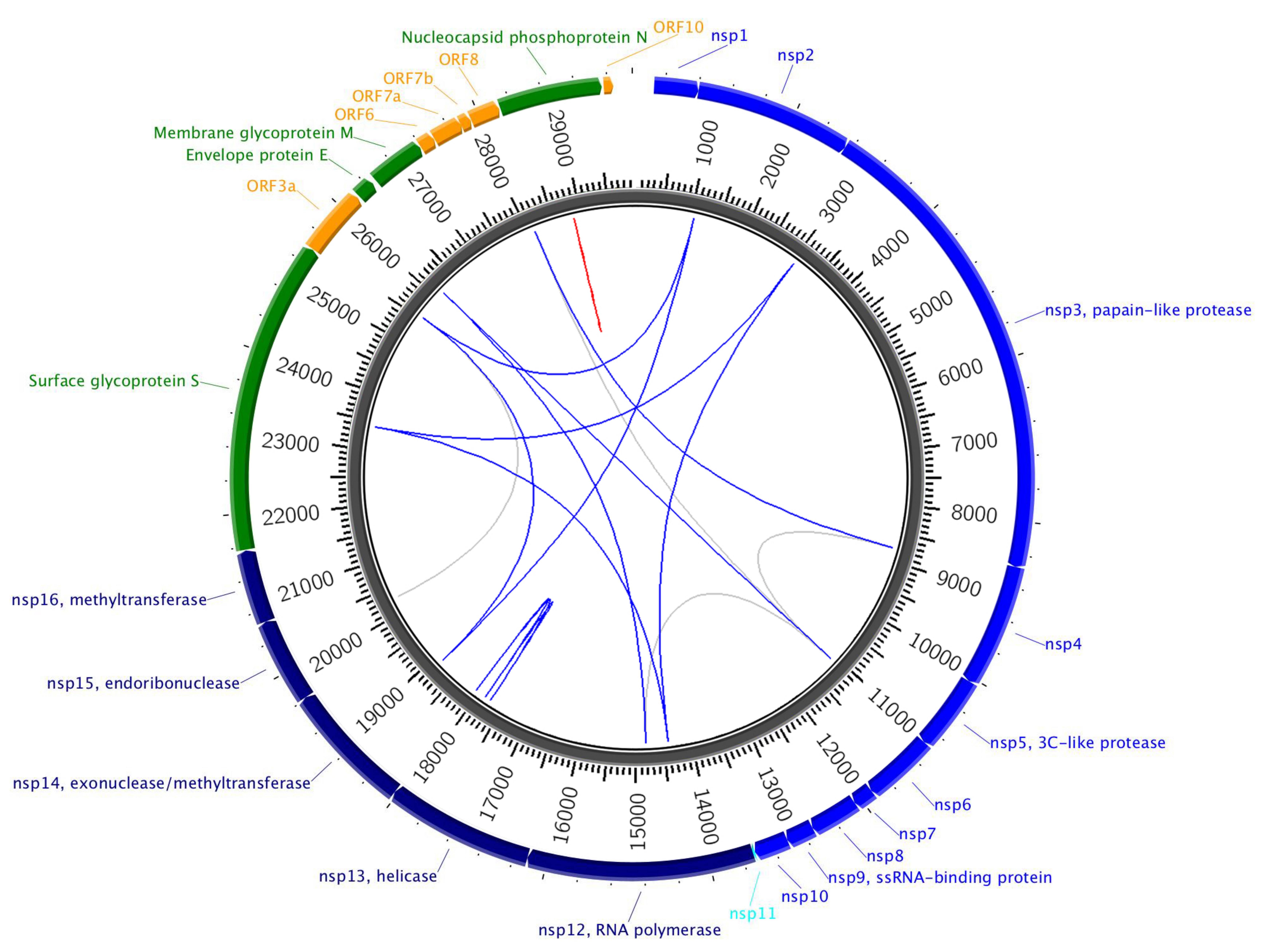}
\caption{Top-200 significant pairwise epistasis from the 2020-08-08 data-set between loci in coding regions. Colored lines indicate for top 50s, grey lines top 50-200. Red lines show short-distance links (distance less than or equal to 3 bp) blue lines show links of longer distance. The colourful links are the same pairs as listed in Table \ref{tab:tab1}. Analogous circos plots for the 2020-05-02 data set is
shown in Appendix \ref{app:20200502}, and for the
2020-04-01 and 2020-04-08 data-sets on the GitHub repository \cite{Zeng-github}.}
\label{fig:circos}
\end{figure}

In Table~\ref{tab:tab1} we list the
significant links
for the 2020-08-08 data-set. The first column is the index of each pairwise interaction in the top 200s. The second column indicates the locus with lower genomic position in the pair and the name of the SARS-CoV-2 proteins it belongs to. The third column lists the major / minor allele (most prevalent, second most prevalent nucleotide) and the mutation type at that locus. The following two columns provide similar information on the
locus with higher
genomic position in the pair.
The last column contains the PLM scores indicating the strength of effects between pairs of loci. The pairwise epistasis listed in Table~\ref{tab:tab1} for 2020-08-08 dataset are visualized by circos software in Fig.~\ref{fig:circos}, where the red ones for the close effects (the distance between two loci is less or equal to 3 locus) while blue for distant effects. Analogous results for the
2020-05-02 dataset is shown in Appendix \ref{app:20200502}, and for the
2020-04-01 and 2020-04-08 data-sets on \cite{Zeng-github}.

\begin{figure}[htpb]
\centering
\includegraphics[width=0.4\textwidth]{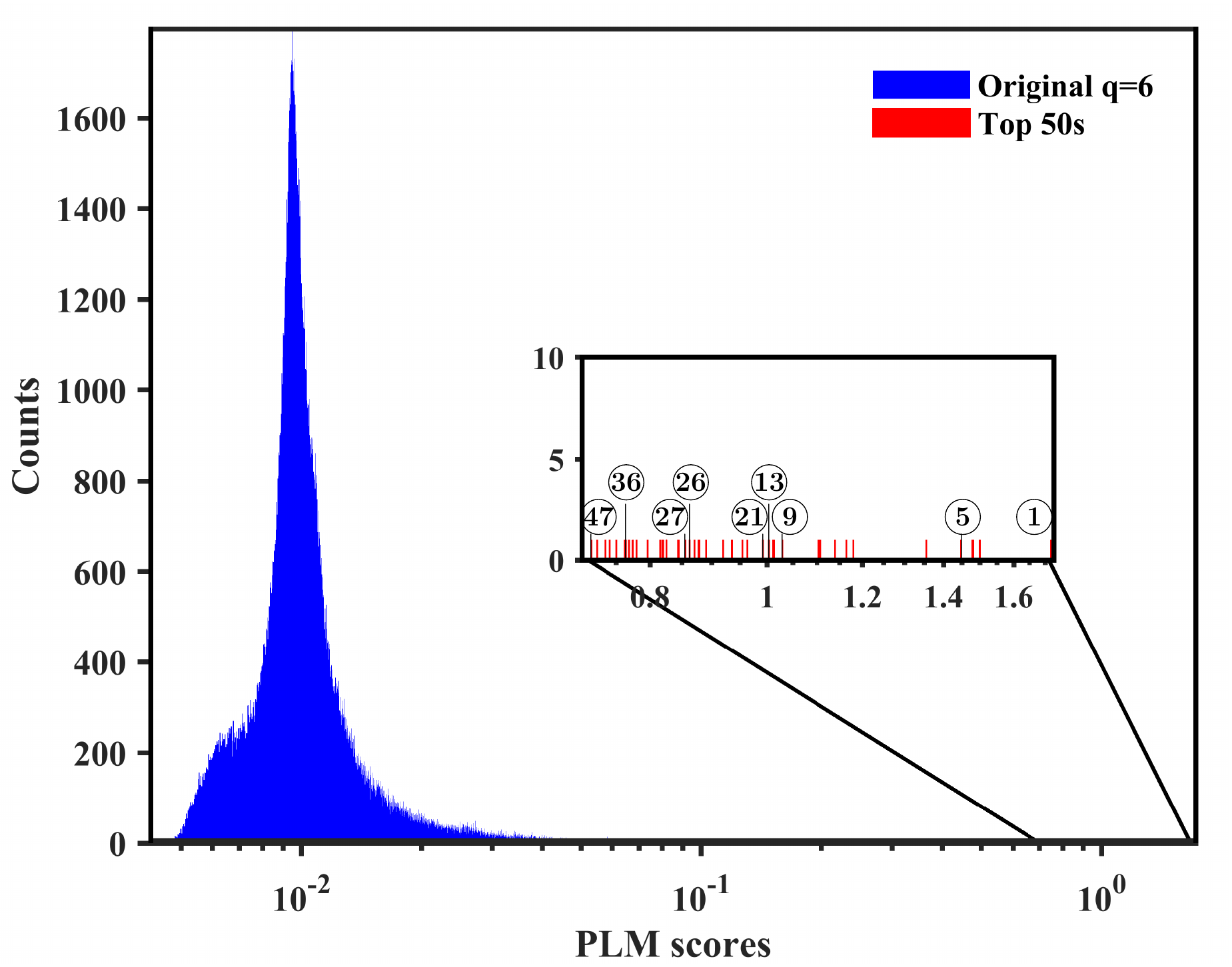}
 \put(-175,135){(a)}\\
\includegraphics[width=0.4\textwidth]{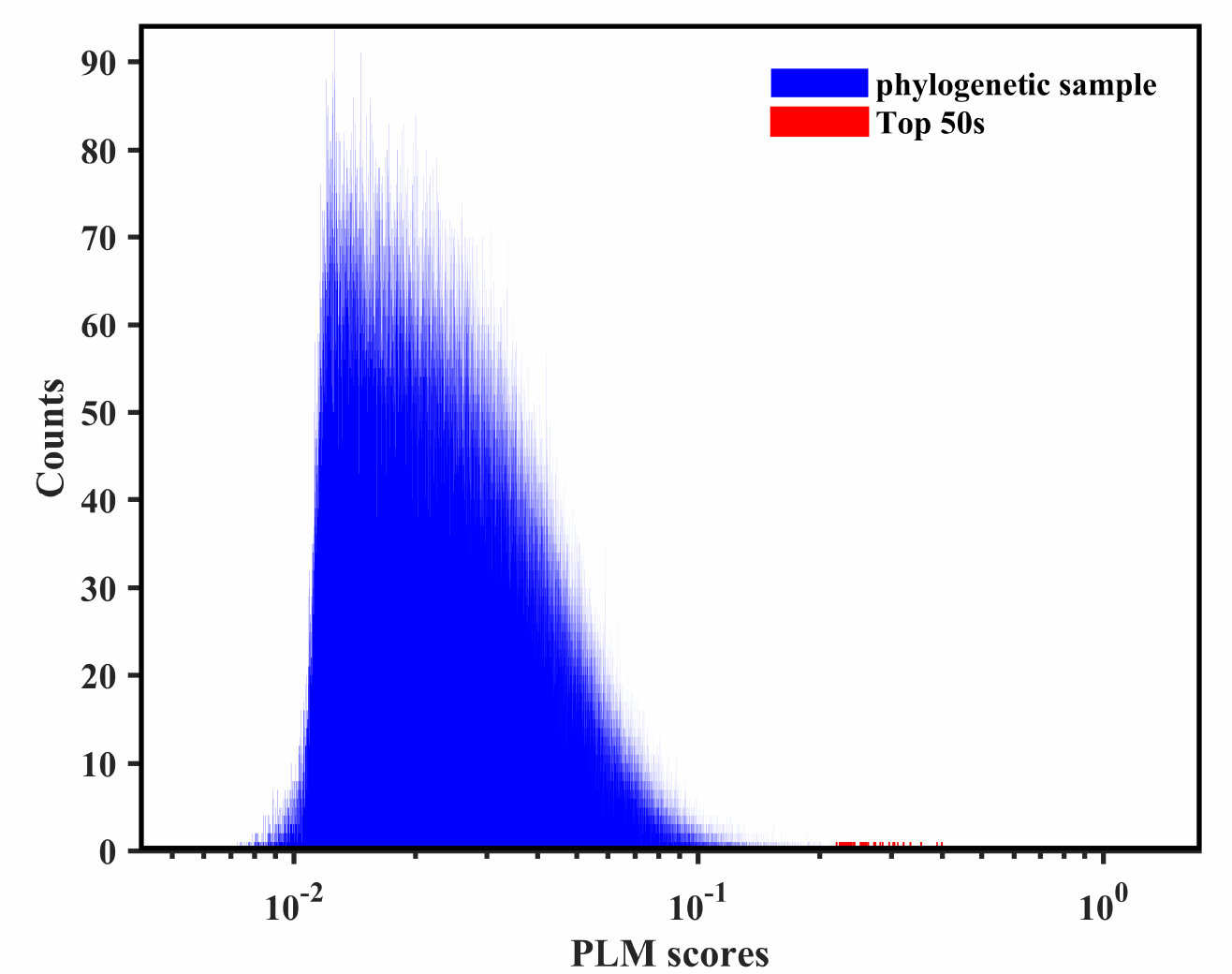}
\put(-175,135){(b)}
{\color{red} \put(-48.5,23){$\downarrow$} }\\
\includegraphics[width=0.4\textwidth]{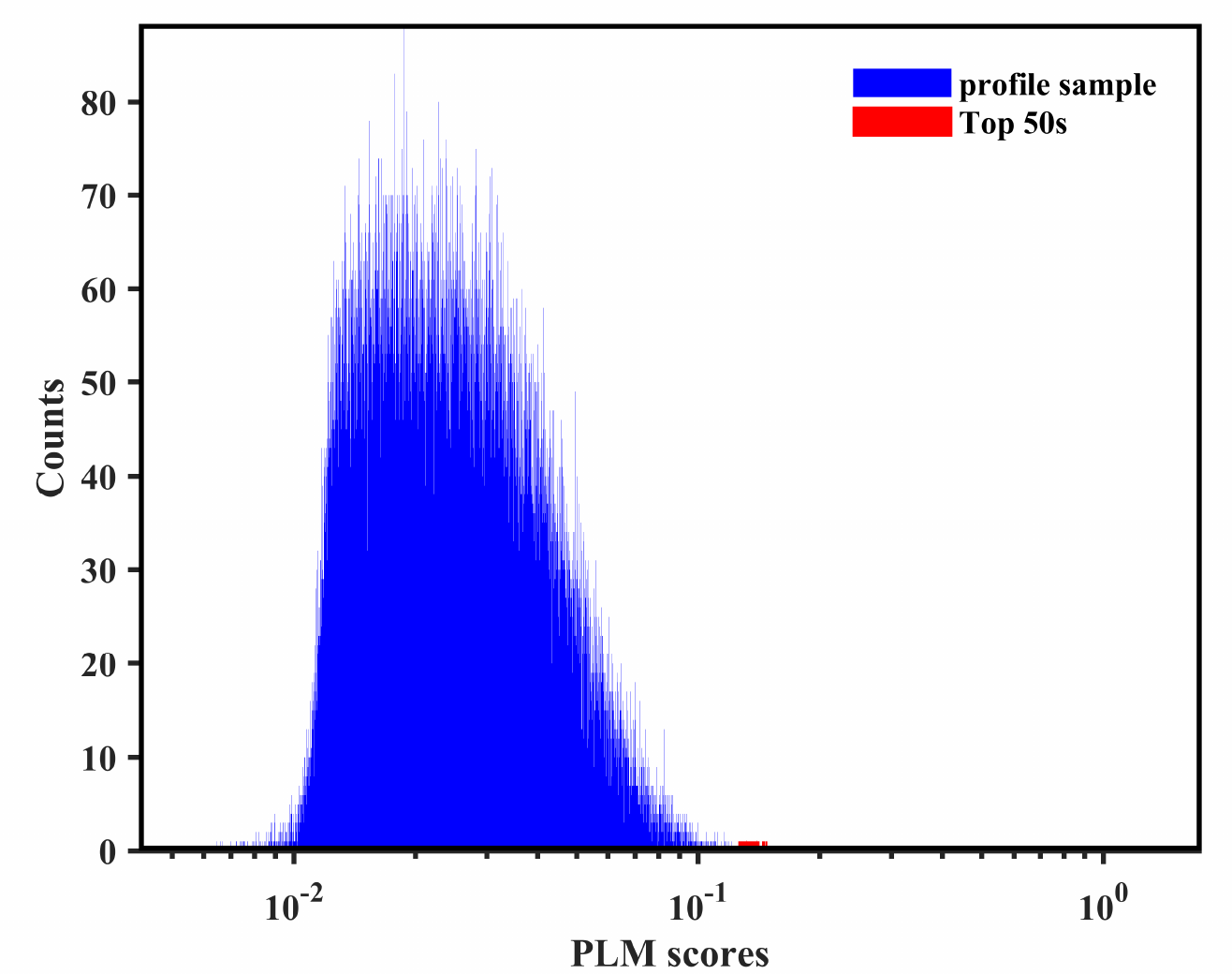}
\put(-175,135){(c)}
{\color{red} \put(-77,23){$\downarrow$}}
\caption{Histograms of PLM scores for (a) original 2020-08-08 data-set,  (b) a phylogenetic randomized sample and (c) a profile randomized sample. The blue bars for all scores while the red ones for top-50 largest scores.
Red arrows in (a) indicate links listed in Table~\ref{tab:tab3}. The largest PLM score is pointed to by red arrows for random samples in (b) and (c). None of them is located inside a coding region, and none of them appear in
Table~\ref{tab:tab1} and Table~\ref{tab:tab3}.}
\label{fig:hists}
\end{figure}

To check if the interactions can be explained by phylogeny
(inherited variations) we used two
randomization strategies
`profile' and `phylogeny' of the Multiple Sequence Alignments (MSAs).
Profile preserves the distribution over alleles at every locus
but does so independently at each locus. Profile
hence destroys all systematic co-variations between loci.
Phylogeny additionally preserves the genetic distance
between each pair of sequences. Viral genealogies inferred
from this information are therefore unchanged under
this randomization.
PLM scores run on these two types of randomized
data (scrambled MSAs) is a background from which the significance of the interactions
from the original data can be assessed.
Each randomization strategy is repeated 50 times with different
realizations of the scrambling, see Appendix \ref{app:phylogenetic_results_visulization}  and~\cite{Zeng-github}.
As shown in Fig.~\ref{fig:hists} the distribution of PLM scores
using phylogeny and profile are
qualitatively different from PLM scores of the
original MSA, with progressively fewer interactions at high score values.
With profile randomization, no interactions predicted by PLM
appear with scores standing out from the background.
Phylogeny randomization on the other hand preserves some interactions found by PLM in a fraction of the realizations of the random background.
Table~\ref{tab:tab2} lists interactions predicted by PLM that appear in some phylogeny
randomizations with scores large compared to the background.
In the following analysis we have not retained them,
see Appendix \ref{app:phylogenetic_results_visulization}
for circos visualizations.
Table~\ref{tab:tab3} lists the eight interactions found by PLM which either do not appear in
any phylogeny randomization with scores that stand out from the background,
or, in the case of (1059-25563), shows up
three times in top-200 out of 50 samples.
We retain these eight
predicted epistatic interactions in the sampled populations of SARS-CoV-2 genomes. %
The top ones listed in Table~\ref{tab:tab3} are marked by red arrows in Fig.~\ref{fig:hists}(a).

\begin{figure}[htpb]
\centering
\includegraphics[width=0.4\textwidth]{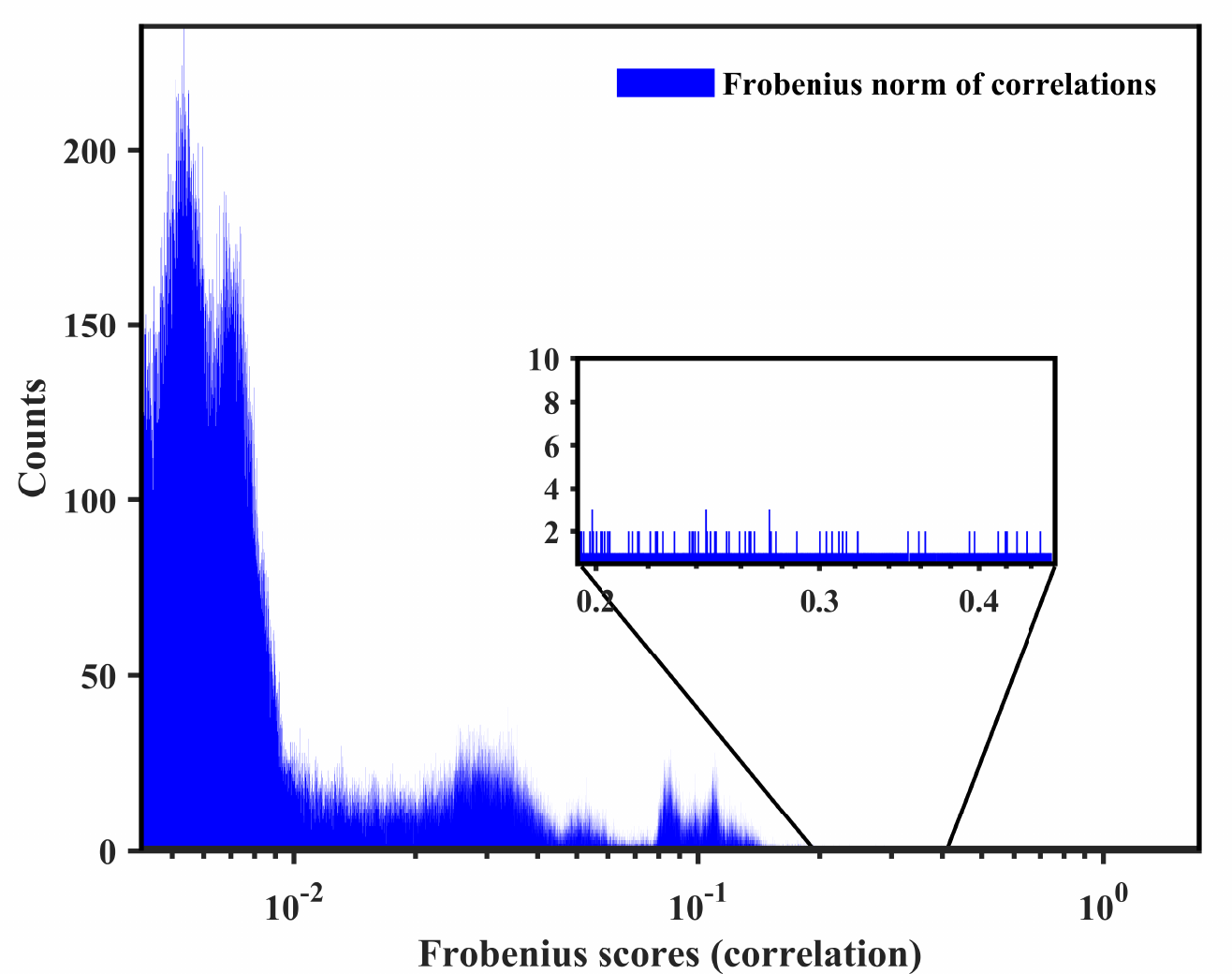}
{\color{red}\put(-75.5,74){\Large{$\downarrow$}}}
\caption{Frobenius norm of pairwise correlations between loci for the original 2020-08-08 data-set. The score pointed by the red arrow corresponds to the link of 1059-25563.}
\label{fig:corr-score}
\end{figure}

Epistatic interactions obtained from DCA reflect
pairwise statistical associations, but not
correlations.
As reviewed in~\cite{Nguyen2017}, and described in Appendix \ref{app:DCA_methods}, DCA is based on a global probabilistic
model, and therefore ranks inter-dependency
differently than correlations.
Fig.~\ref{fig:corr-score} compared to
Fig.~\ref{fig:hists} shows that the distribution of
correlation scores is qualitatively different
from the distributions of DCA scores in the GISAID data set.
Fig.~\ref{fig:corr-PLM-score-decrease} further
shows that the rank of the
epistatic interaction predicted in Table~\ref{tab:tab3}
have remained stable, while the corresponding correlations
have merged into the background.

The first-ranked interaction between 1059 and 25563 is between a
(C/T), resulting in the T85I non-synonymous mutation in gene nsp2 and
a (G/T), resulting in the Q57H non-synonymous mutation in gene ORF3a.
nsp2, expressed as part of the ORF1a polyprotein,
binds to host proteins prohibitin 1 and prohibitin 2 (PHB1 and PHB2) in SARS-CoV~\cite{Yoshimoto2020}.
The variations in the site 1059 have been predicted
to modify nsp2 RNA secondary structure~\cite{Rad2020implications}
and have previously been reported to co-occur together with the Q57H variant in ORF3a in a dataset of SARS-CoV-2 genomes from the United States~~\cite{Wang2020}.
ORF3a, also known as ExoN1 hypothetical protein sars3a, forms a cation channel
of which the structure in SARS-CoV-2 is known by Cryo-EM~\cite{Kern-2020}.
In SARS-CoV ORF3a been shown to up-regulate expression of fibrinogen subunits FGA, FGB and FGG in host lung epithelial cells~\cite{Tan2005},
to form an ion channel which modulates virus release~\cite{Lu2006},
to activate the NLRP3 inflammasome~\cite{Siu2019},
and has been found to induce apoptosis~\cite{Ren2020}. The Q57H variant was reported early in the COVID-19 pandemic~\cite{Issa2020}
and occurs in the first transmembrane alpha helix, TM1~\cite{Kern-2020}, where it changes the amino acid glutamine (Q) with a non-charged polar side chain to histidine (H), which has a positively charged polar side chain. This amino acid is at the interface of interaction between the two dimeric subunits of ORF3a that forms the constrictions of the ion channel but the Q57H alteration does not seem to change the ion channel properties compared to wildtype 3a ~\cite{Kern-2020}. Nevertheless, its incidence is increasing in SARS-CoV-2 genomes in the United States~\cite{Wang2020}
and the effect of Q57H may therefore affect the virulence in other beneficial ways than changing the conductance properties of the ion pore.

The association between 8782 and 28144
(rank 5), reported early in SARS-CoV-2 studies~\cite{Tang2020}
is between a (C/T) synonymous mutation in the gene
nsp4, and a (T/C) non-synonymous mutation resulting in the L84S alteration in the gene ORF8.
The first of these genes participates in the assembly of virally-induced cytoplasmic double-membrane vesicles necessary for viral replication.
The site 8782 is located in a region annotated as
CpG-rich and is the site of a CpG for the major allele (C);
it has the minor (T) allele in other related viruses~\cite{Tang2020}. Orf8 has been implicated in regulating the immune response~\cite{Zhang2020orf8,Li2020orf6}.
The L84S variant is, together with the C8782T nsp4 mutation characterizing the GISAID clade S~\cite{Mercatelli2020geographic}.

The interaction between 14805 and 26144 (rank 9) leads to non-synonymous alterations in nsp12 (T455I, note that the reference is Y) and ORF3a	(G251V) respectively. The G251V has been reported by many studies and is defining the GISAID V clade ~\cite{Mercatelli2020geographic} together with the L37F nsp6 variant (position 11083, rank 47). The widely reported G251V variant is unfortunately outside of the proposed Cryo-EM structure ~\cite{Kern-2020} and it is unknown how this glycine to valine substitution affects protein function.
nsp12 is the RNA-dependent RNA polymerase and the T455I substitution is found where the reference Wuhan-Hu-1 has a tyrosine residue in one of the alpha helices of the polymerase "finger" domain~\cite{Chen2020}.
Threonine can similarly to tyrosine be phosphorylated but also glycosylated, it is polar, uncharged and can form hydrogen bonds that may stabilize the alpha helix. Isoleucine on the other hand is non-polar and uncharged and both the residues are smaller than the aromatic tyrosine.

The second interaction partner of G251V is the nsp6 L37F variant. nsp6 has been shown to induce autophagosomes in the host cells in favour for viral replication and propagation SARS-CoV~\cite{Yoshimoto2020}. There is currently no experimentally validated model of nsp6 structure but an early model suggest that the L37P variant is situated in an unordered loop between two alpha helices~\cite{Benvenuto2020}.

The interaction between 17747 and 17858 (rank 27) is between
two non-synonymous mutations (C/T, resulting in P504L) and (A/G, resulting in T541C)  within
the gene nsp13 that codes for a
helicase enzyme that unwinds duplex RNA~\cite{Yoshimoto2020}.
It is the only epistatic interaction in Table~\ref{tab:tab3}
within one protein.
These same two loci reappear in the list with ranks 26 and 36 as interacting with a C/T synonymous mutation (L7L) in gene nsp14 at position 18060. The P504L and T541C are both located in the Rec2A part of the protein that is not in direct interaction with the other members of the  RNA-dependent RNA polymerase holoenzyme, in which two molecules of nsp13 forms a stable complex with nsp12 replicase, nsp7, and nsp8. The nsp14 protein is a bifunctional protein that has a N7-methyltransferase domain and an a domain exonuclease activity, responsible for replication proof reading (cite Denison et al "An RNA proofreading machine regulates replication fidelity and diversity"). The nsp14/nsp10 proof reading machinery is thought to interact with the replication-transcription complex but the exact details of this interaction are not known.

The final interaction (rank 21) is a link between a locus carrying a non-synonymous
mutation (C/T, T541C) in nsp2 position 1059, with
a locus carrying a synonymous mutation (C/T, L280L) in nsp14, position 18877. As the knowledge on nsp2 protein structure is poor there is no evidence for the effect of this mutation. Also, how the synonymous C/T alterations in nsp14, as well as in the synonymous mutations of the other interactions affect the virus are unknown, but can be proposed to change RNA secondary structure, RNA modification or codon usage.

\section{Discussion}\label{sec:Discussion}
The COVID-19 pandemic is a world-wide public health emergency caused by the $\beta$-coronavirus SARS-CoV-2. A very large and continuously increasing number of  high-quality whole-genome sequences are available. We have investigated whether these sequences  show effects of epistatic contributions to fitness. In a population evolving under high rate of recombination, such effects of natural selection can be detected by Direct Coupling Analysis, a global model learning technique. The paper opens up the prospect to leverage very large collections of genome sequences to find new combinatorial weaknesses of highly recombinant pathogens.

In this work we have considered all whole-genome sequences of SARS-CoV-2
deposited in GISAID up to different cut-off dates. As this coronavirus
has extensive recombination we have assumed that the distribution
of genotypes is well described by Kimura's Quasi-Linkage Equilibrium,
and used Direct Coupling Analysis to infer epistatic contributions to
fitness from the sequences.
After filtering out all but the strongest effects and variations in non-coding regions with many gaps in the
MSA, the remaining predictions are
few in number, \textit{i.e.} \textbf{19} predictions in Table~\ref{tab:tab1}.

Co-variations between allele distributions at different loci can be due
to epistasis and also to inherited effects (phylogeny). We have tested for the
second type by randomizing Multiple Sequence Alignment of sequences
such that pair-wise distances between sequences are left invariant.
We find that the top link 1059-25563 appears 3 times in 50 phylogeny randomizing samples, though with much lower rank. The other
predicted epistatic contributions disappear under phylogenetic randomization,
except for pairs in the triple (3037, 14408, 23403) which appear in from 20\% to 35\% of 50
randomizations. After eliminating these links as well as links
between adjacent
loci (28881, 28882, 28883, which appear in from 14\% to
16\% in 50 samples), we are left with eight predictions listed
in Table~\ref{tab:tab3}.
We consider it likely that these
retained interactions are due to epistasis, and not to
inherited co-variation.
An analogous investigation on a smaller dataset obtained
with an earlier cut-off date (2020-05-02) and reported in Appendix \ref{app:20200502} yield six retained predictions, involving
however the same eight viral genes.
The question on epistasis vs. effects of inheritance (phylogeny) clearly
merits further investigation and
testing as more data will become available.

Biological fitness is a many-sided concept and
can also include aspects of game and cooperation~\cite{MaynardSmith82,Nowak2004,ClaussenTraulsen2008}.
A fitness landscape describes the propensity of an individual
to propagate its genotype in the absence of strategic interactions
with other genotypes, and has traditionally been used to
model the evolution of pathogens colonizing a
host, for earlier use relating to HIV and
using DCA techniques, see~\cite{FERGUSON2013}.
The additive and epistatic contributions to fitness of the
virus which we find describe the virus in its human host
and therefore likely reflect host-pathogen interactions to a large extent.
A conceptual simplification made is that all hosts have been assumed equivalent. In
future methodological studies it would be of interest to consider
possible effects of evolution in a collection of
landscapes, representing different hosts,
and to correlate such dynamics to host genotypes.
As this requires other data than available on GISAID, and
less abundant at this time, we leave this for future work.
On the other hand, it is unlikely that the inferred
couplings involve the host as a temporal variable, due to the much
faster time scale of the evolution of the virus.

\begin{figure}[htpb]
\centering
\includegraphics[width=0.45\textwidth]{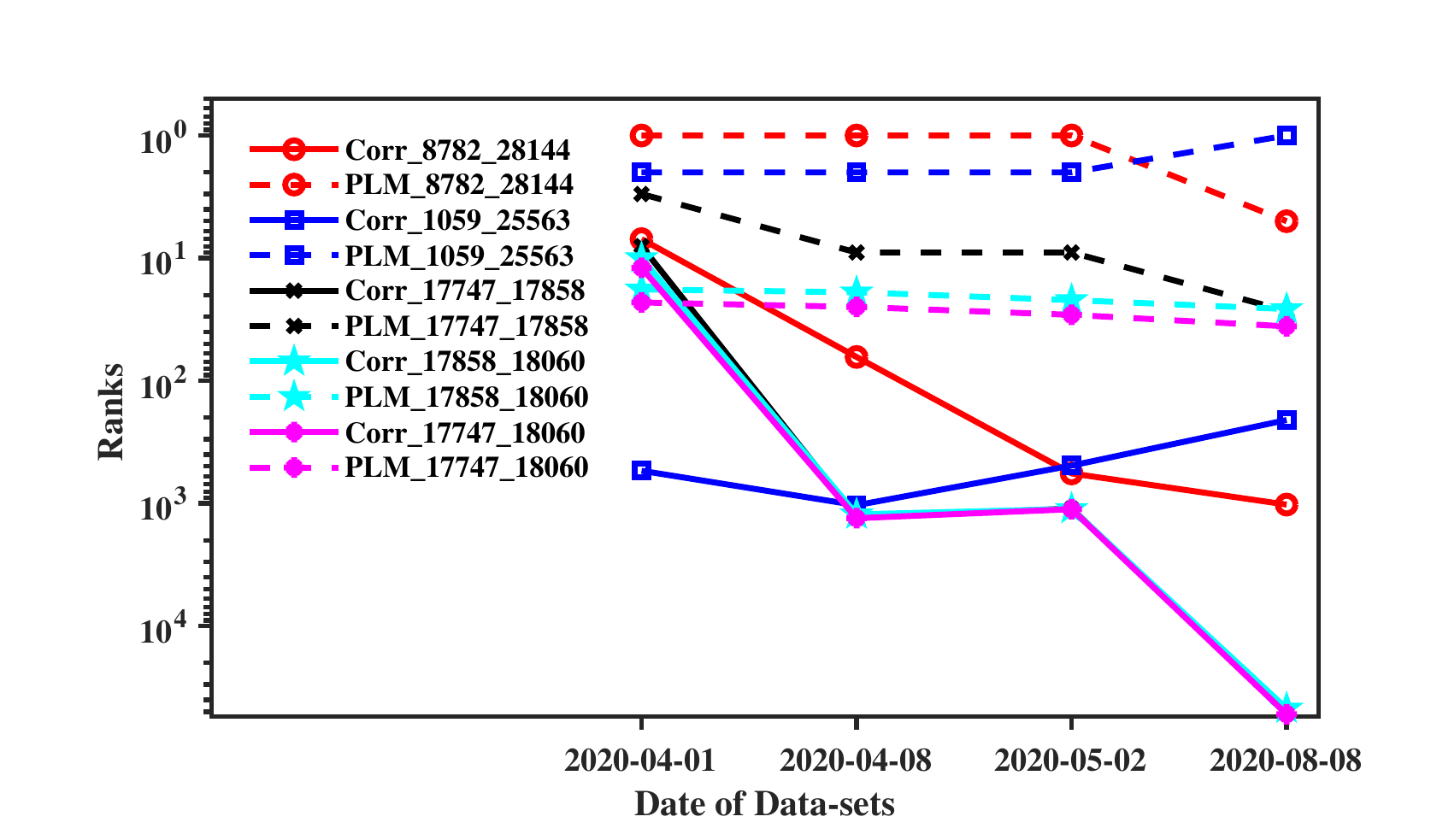}
\caption{Ranks for significant epistatic effects with data collection date (2020-04-01, 2020-04-08, 2020-05-02 and 2020-08-08) by PLM (dashed lines) and correlation analysis (solid lines). The ranks of the PLM scores are almost constant while the ranks of the correlations
vary significantly and mostly drop as more data accumulate (later cut-off dates).}
\label{fig:corr-PLM-score-decrease}
\end{figure}

Epistatic interactions are pairwise statistical associations,
but are not simply correlations. The interaction
between sites 8782 and 28144, which is the second largest in
Table~\ref{tab:tab3},
was identified as a very strong correlation in
a very early study~\cite{Tang2020}.
As shown in Table~\ref{fig:corr-PLM-score-decrease}
this correlation has generally decreased over time
(using data with successively later cut-off dates).
In the alternative
global model learning method of DCA
which we use in the present work, the score of
statistical inter-dependency of this pair
has remained large, and the pair is consistently
ranked first or second over four different cut-off dates, see
Fig.~\ref{fig:corr-PLM-score-decrease}.
While our data hence supports the observation
of statistical inter-dependency in this pair first made in~\cite{Tang2020},
it does not support the interpretation made in the same
work that the effect is due to phylogeny.
The later criticism in~\cite{MacLean2020}
therefore does not apply to our work since an
epistatic interaction, recovered through DCA
and a Quasi-Linkage Equilibrium assumption
in a population thoroughly mixed by recombination,
is different in nature from a phylogenetic effect.

DCA techniques have been applied to find candidate targets
for vaccine development.
In a series of studies it was found that
combinations of mutations implied by sequence variability
in the HIV-1 Gag protein correlate well with \textit{in vitro}
fitness measurements, and clinical observations
on escape strains (HIV strains that tend to dominate in one patient over time)
and the immune system of
elite controllers (HIV-positive individuals progressing slowly towards
AIDS)~\cite{Dahirel2011,FERGUSON2013,Mann2014}.
While this may be a promising future avenue in COVID-19 research,
in the present study we have not found any epistatic interactions
involving Spike, only pairs that also show up under phylogeny
randomization or that are quite weak, see Appendix \ref{app:D614G}.
The Spike protein has been the main target
of coronovirus vaccine development to date~\cite{Tse2020},
including against
SARS-CoV-2~\cite{Amanat2020,Amanat2020,TungThanLe-2020-1,TungThanLe-2020-2}.


An epistatic interaction means that loss of fitness by a mutation
at one locus is
enhanced (positive epistasis) or compensated (sign epistasis) by a mutation
at another locus. Suppose there are drugs that act on targets around both
loci, modulating the fitness of the respective variants. Epistasis then
points to the possibility that
using both drugs simultaneously may have a more than additive effect.
To search if our analysis offers such a guide to
combinatorial drug treatment,
we scanned the recent comprehensive compilation of drugs
known or predicted to target SARS-CoV-2~\cite{Gordon2020}.
Five out of the eight predictions in Table~\ref{tab:tab3}
involve either one synonymous mutations or are between
two mutations in the same gene.
For all the three remaining pairs of non-synonymous
mutations, (1059, 25563), (11083, 26144)
and (14805, 26144), the second locus
lies in ORF3a for which
no potential drugs are listed in~\cite{Gordon2020}.
The first locus in the same three pairs lie
respectively in genes nsp2, nsp6 and nsp12.
One or more already approved and practical drugs
targeting nsp2 and nsp6 are
listed in~\cite{Gordon2020}.
Ponatinib, listed for nsp12,
is not appropriate against a
pandemic disease like COVID-19 on account of its large cost. Potential drugs for the proteins listed in Table~\ref{tab:tab3} are summarized in Table \ref{tab:drugs} in Appendix \ref{app:potential_drags}, following ~\cite{Gordon2020}.

Nevertheless, the number of combinations of potential drug targets,
in COVID-19
and many other diseases, is very large. Direct Coupling Analysis
applied to many sampled sequences predicts which genes/loci
have mutual dependencies in fitness, and can be used to
rank combinations for further more detailed investigation.
We note that one can also start a search for drug treatment
from conserved positions, assuming these to be unconditionally
necessary for the virus. If so, all potential pairs would however
be ranked equal based on sample information,
and there would be no analogous short-cut to the combinatorial explosion of possibilities.
Even if the scan discussed above did not lead to any direct suggestions
based on the lists of potential drugs in~\cite{Gordon2020}, we hope
the general approach could have value given the continuing increase and
availability of genome sequences of both viral and bacterial pathogens.
We finally note three out of eight of our list of predictions
involve loci in viral gene ORF3a, the action of which is related to
severe manifestations of COVID-19 disease~\cite{Lu2006,Siu2019,Ren2020}.

\begin{table*}[htpb]
\begin{ruledtabular}
\caption{Significant links with rank within top-200s between pairwise loci for the 2020-08-08 data-set.}
\begin{tabular}{llllll}
\textrm{Rank\footnote{Indices of significant links in the top 50s with both terminals located inside a coding region, inferred by PLM. The analogous table for the 2020-05-02 data-set is shown in Appendix \ref{app:20200502}}} & \textrm{Locus 1\footnote{Information on locus 1: index in the reference sequence, the protein it belongs to. The convention used is that locus 1 (``starting locus")
is the locus of lowest genomic position in the pair.}}  &  \textrm{mutation\footnote{Information on mutations of locus 1: the first and second prevalent nucleotide at this locus, mutation type: synonymous(syn.) / non-synonymous(non.).}} & Locus 2 & mutation & PLM   \\
~   & -protein & -type & -protein & -type & score  \\
\hline
1&1059-nsp2 & C$|$T-non. & 25563-ORF3a & G$|$T-non. & 1.7191 \\
2&28882-N & G$|$A-syn. & 28883-N & G$|$C-non. & 1.4996 \\
3&28881-N & G$|$A-non. & 28882-N & G$|$A-syn. & 1.4816 \\
4&28881-N & G$|$A-non. & 28883-N & G$|$C-non. & 1.4783 \\
5&8782-nsp4 & C$|$T-syn. & 28144-ORF8 & T$|$C-non. & 1.4471 \\
9&14805-nsp12 & C$|$T-syn. & 26144-ORF3a & G$|$T-non. & 1.1392 \\
12&3037-nsp3 & T$|$C-syn. & 14408-nsp12 & T$|$C-non. & 1.0291 \\
13&18877-nsp14 & C$|$T-syn. & 25563-ORF3a & G$|$T-non. & 1.0131 \\
14&3037-nsp3 & T$|$C-syn. & 23403-S & G$|$A-non. & 1.0114 \\
17&14408-nsp12 & T$|$C-non. & 23403-S & G$|$A-non. & 0.9917 \\
21&1059-nsp2 & C$|$T-non. & 18877-nsp14 & C$|$T-syn. & 0.9197 \\
26&17858-nsp13 & A$|$G-non. & 18060-nsp14 & C$|$T-syn. & 0.8624 \\
27&17747-nsp13 & C$|$T-non. & 17858-nsp13 & A$|$G-non. & 0.8553 \\
36&17747-nsp13 & C$|$T-non. & 18060-nsp14 & C$|$T-syn. & 0.7780 \\
47&11083-nsp6 & G$|$T-non. & 26144-ORF3a & G$|$T-non. & 0.7340 \\
63&20268-nsp15 & A$|$G-syn. & 25563-ORF3a & G$|$T-non. & 0.6474 \\
134&11083-nsp6 & G$|$T-non. & 14805-nsp12 & C$|$T-syn. & 0.5040 \\
147&11083-nsp6 & G$|$T-non. & 28144-ORF8 & T$|$C-non. & 0.4928 \\
168&8782-nsp4 & C$|$T-syn. &11083-nsp6 & G$|$T-non.  & 0.4770 \\
\end{tabular}
\label{tab:tab1}
\end{ruledtabular}
\end{table*}

\begin{table}[htpb]
\caption{Top 200s that \textbf{appeared} (with an appearance ratio $\ge 10\%$ ) in samples with phylogeny randomization strategy based on the 2020-08-08 data-set. \textbf{50} phylogeny samples are considered in total.
}
\begin{ruledtabular}
\begin{tabular}{llllll}
\textrm{{Hit} \footnote{The indices of samples with phylogeny randomization which preserve the links listed in Table~\ref{tab:tab1} are shown here. The circos plots for the significant epistatic links of all \textbf{50} randomized samples are available in SI}} & Locus 1 & mutation & Locus 2 & mutation  \\
ratio & -protein & -type & -protein & -type \\
\hline
14\%  & 28881-N & G$|$A-non. & 28882-N & G$|$A-syn. \\
16\%  & 28881-N & G$|$A-non. & 28883-N & G$|$C-non. \\
20\%  & 28882-N & G$|$A-syn. & 28883-N & G$|$C-non. \\
22\%  &3037-nsp3 & T$|$C-syn. & 14408-nsp12 & T$|$C-non.\\
20\%&3037-nsp3 & T$|$C-syn. & \textrm{23403-S \footnote{ In amino acid notation this mutation is D614G in Spike.}}  & G$|$A-non. \\
34\% &14408-nsp12 & T$|$C-non. &  23403-S  & G$|$A-non.\\
\end{tabular}
\end{ruledtabular}
\label{tab:tab2}
\end{table}

\begin{table}[htpb]
\caption{Potentially significant epistatic links in Table \ref{tab:tab1}, and corresponding amino acid mutations}
\begin{ruledtabular}
\begin{tabular}{lllll}
\textrm{Rank\footnote{Main prediction: eight epistatic links. The links preserved by phylogeny randomization in Table \ref{tab:tab2} are not listed here.}} & Locus 1- & amino acid & Locus 2- & amino acid \\
~&  protein  & mutation & protein & mutation \\
\hline
\textrm{1\footnote{This link appears in $3$ out of $50$ ($6\%$)
phylogeny randomizations; once (experiment 23)
with rank 34, and twice (experiments 29 and 47)
with ranks in $51-200$, see Appendix \ref{app:phylogenetic_results_visulization}.}} & 1059-nsp2 & \textrm{T85I(T\footnote{Amino acid in the reference sequence Wuhan-Hu-1 at the position specified by the number between major and minor alleles.})} & 25563-ORF3a  & Q57H(Q)  \\
5& 8782-nsp4 & S76S(S) & 28144-ORF8 & L84S(L) \\

9&14805-nsp12 & T455I(Y) & 26144-ORF3a  & G251V(G) \\

21 &1059-nsp2 & T85I(T) & 18877-nsp14  & L280L(L) \\

26&17858-nsp13  & T541C(Y) & 18060-nsp14 & L7L(L) \\

27&17747-nsp13 & P504L(P) & 17858-nsp13  & T541C(Y) \\

36&17747-nsp13 & P504L(P) & 18060-nsp14 & L7L(L)\\

47&11083-nsp6 & L37F(L) & 26144-ORF3a & G251V(G) \\
\end{tabular}
\end{ruledtabular}
\label{tab:tab3}
\end{table}

\begin{table}[htpb]
\caption{Top-10 links found by correlation analysis in the coding region for the data-set till 2020-08-08.}
\centering
\begin{ruledtabular}
\begin{tabular}{lllc}
\textrm{Rank\footnote{Rank for top-10 links as ranked by correlation analysis. Correlations between loci of which at least one outside coding regions are omitted.}} & Locus 1  & Locus 2 & Frobenius  \\
~ & -protein & -protein  & Score\\
\hline
455 &  3037-nsp3   &   23403-S &	0.3844\\
458 &	 3037-nsp3   &   14408-nsp12  & 0.3842\\
460 &  14408-nsp12 &	23403-S &	0.3837\\
581 & 28882-N   & 28883-N  & 0.3609 \\
584 & 28881-N  & 28883-N  & 0.3603 \\
585 & 28881-N & 28882-N & 0.3603\\
1071 & 1059-nsp2 & 25563-ORF3a & 0.2821\\
2394 & 8782-nsp4 & 28144-ORF8 & 0.1803\\
3969 & 23403-S & 28144-ORF8 & 0.1487\\
3980 & 3037-nsp3 & 28144-ORF8 & 0.1486\\
\end{tabular}\\
\end{ruledtabular}
\label{tab:tab4}
\end{table}

\begin{acknowledgments}
We thank Profs Martin Weigt and Roberto Mulet for numerous discussions, and
Martin Weigt for sharing the use
of the `phylogeny' developed with ERH.
EA thanks 
Arne Elofsson and other participants
of the Science for Life Labs (Solna, Sweden) "Viral sequence evolution research program" for input and suggestions.
The work of HLZ was sponsored by National Natural Science Foundation
of China (11705097), Natural Science Foundation of Jiangsu Province (BK20170895),
Jiangsu Government Scholarship for Overseas Studies of 2018, and Scientific Research
Foundation of Nanjing University of Posts and Telecommunications (NY217013).
The work of Vito Dichio was supported by Extra-Erasmus Scholarship (University of Trieste) and Collegio Universitario 'Luciano Fonda'. ERH acknowledges funding by the
EU H2020 research and innovation programme MSCARISE-2016 under grant agreement No. 734439 INFERNET.
\end{acknowledgments}

\appendix

\section{Quasi-linkage equilibrium (QLE) and its range of validity}\label{app:QLE}

Quasi-linkage equilibrium was discovered by Kimura \cite{Kimura1956} and
investigated more recently by Neher and Shraiman~\cite{NeherShraiman2009, NeherShraiman2011}.
The connection to inference was made
in~\cite{Gao2019} where the theory was also extended
from Boolean variables (bi$-$allelic loci) to
categorical data (arbitrary number of alleles per locus).

QLE refers to the state of a population comprising $N$
individuals. That can be characterized by
the list of genotypes present,
or equivalently by an
\textit{empirical probability distribution}
\begin{equation}
\label{eq:empirical}
P^{(e)}(\mathbf{\sigma})=\frac{1}{N}\sum_s \mathbf{1}_{\mathbf{\sigma},\mathbf{\sigma}^{(s)}}
\end{equation}
In above  $\mathbf{\sigma}^{(s)}$ is the genotype of
individual $(s)$
and $\mathbf{1}_{\mathbf{\sigma},\mathbf{\sigma}^{(s)}}$
means that individuals of genotype $\mathbf{\sigma}$ are counted in the sum.
Each genotype can hence appear zero, once or many times;
if more than zero such a group is referred to as a \textit{clone},
and if more than once a multi-individual clone, or a proper clone.
We will in the following for simplicity take $N$ to be fixed.
In QLE there are few proper clones,
i.e. most individuals in the population have a unique genotype.

In models of evolution where the QLE concept is pertinent
populations change in time due to mutations, recombination, fitness
and genetic drift.
Mutation and recombination are random events characterized
by average rates. Fitness is the propensity of an individual to
have offspring in the next generation and genetic drift is
the randomness associated with this process.
The population history is hence given by
a sequence of empirical probability distributions
indexed by time ($t$), $P^{(e)}(\mathbf{\sigma},t)$.

The development in time can alternatively be formulated on
the level of \textit{ensemble probability distributions}. In general that
will be the
time-dependent joint probability distributions of $N$ genomes
$P_N(\mathbf{\sigma}_1,\mathbf{\sigma}_1,\ldots,\mathbf{\sigma}_N,t)$.
Mutation, fitness and genetic drift act on each genotype
in the population separately.
They can thus be formulated for the
marginalization $P_N$ to a one-genome
ensemble probability distribution $P(\mathbf{\sigma},t)$.
Recombination acts on two genomes at a time.
They can thus be formulated on the level of
two-genome
ensemble probability distributions $P_2(\mathbf{\sigma}_1,\mathbf{\sigma}_2,t)$.

QLE is characterized by multi-genome
distribution functions factorizing \textit{i.e.}
by
\begin{equation}
   P(\mathbf{\sigma}_1,...,\mathbf{\sigma}_N)\approx P(\mathbf{\sigma}_1) ... P(\mathbf{\sigma}_N).
\end{equation}
As a consequence, evolution can be
considered on the level of one-genome
ensemble probability distributions only.
The resulting equations have been written out
and discussed in~\cite{NeherShraiman2009,NeherShraiman2011,Gao2019,Zeng2020}
and are structurally similar to the Boltzmann equation in
gas kinetics, where recombination plays the role of a collision term.
It is a reasonable picture to consider the list of $N$ genotypes
given by $P^{(e)}$ as $N$ samples from $P$. However,
as follows from above, the relation between the two distributions
is not direct: $P^{(e)}$ changes
according to a stochastic and frequency-independent evolution law
while $P$ obeys a deterministic nonlinear partial differential equation.

The form of QLE, for only bi-allelic loci and only
one-locus fitness terms (additive fitness) and two-loci fitness terms
(epistatic fitness) is that of the Ising model of
statistical mechanics \textit{i.e.} for a genome
$\mathbf{\sigma}=(s_1,\ldots,s_L)$
\begin{equation}\label{eq:Ising}
P(\mathbf{\sigma})=\frac{1}{Z}\exp\left(\sum_i h_i s_i + \sum_{ij} J_{ij}s_is_j\right)
\end{equation}
In above both sums go over the loci of variability on the
genome. In some applications they could be all or almost all
positions, but in the application to SARS-CoV-2 genomes
they are but a small fraction. At most genomic positions
all samples carry the same variant; those positions are not included
in the sums in \eqref{eq:Ising} which focus on the variable portions.
Further, $s_i=1$ in \eqref{eq:Ising} encodes one of the alleles (typically the major allele)
and  $s_i=-1$ encodes the other allele (typically the minor allele),
$h_i$ and $J_{ij}$ parametrize the distribution
and $Z$ is a normalization constant (partition function).
The extension of the above to more the bi-allelic loci ("Potts genomes")
is given in~\cite{Gao2019},
and restated in main body of the paper, Eqs.~(3) and~(4).
Although we in this work only find properties
of almost bi-allelic pairs of loci, the intermediate
analysis allows multi-allelic states (see below).

Known \textit{sufficient conditions} for QLE
are that recombination is sufficiently high everywhere
compared to selection, and mutation rate is non-zero.
The part of the argument concerning recombination
was made on the physical level of rigor in \cite{NeherShraiman2011},
by estimating terms in a Taylor series in inverse recombination
rate.
The part of the argument concerning mutation is
implicit in \cite{NeherShraiman2011}, and based on the observation that
without mutations, the most fit genotype will
eventually take over in a finite population. This means
that mutation must be non-zero, as otherwise QLE will only
be a (possibly long-lived) transient~\cite{Zeng2020}.
On a qualitative level the argument for QLE is analogous
to the accuracy of Boltzmann equation and
the stationary state being of Gibbs-Boltzmann type when collision rate is high enough~\cite{Gao2019}.

Many realistic models encode
that recombination acts more weakly between closely
spaced loci. Taken at face value, the conditions in \cite{NeherShraiman2009,NeherShraiman2011,Gao2019}
would then imply that the conditions for QLE can be at hand
for loci sufficiently far apart on the genome, but not
for closely spaced loci.
A quantitative theory of this effect as it relates to QLE
is at this point lacking. Numerical results in
\cite{NeherShraiman2009,NeherShraiman2011}
showed that the characteristics of QLE may be present, and
numerical results in~\cite{Zeng2020} showed that QLE can
be used for inference in this setting.

\textit{Necessary conditions} for the validity of QLE
are at this point not known.
It may be that states similar to QLE can also be found
in other parameter ranges than at high recombination.
An argument in favor is the recent in silico result in~\cite{Zeng-Mauri-2020}
where a modified form of QLE inference was shown to work also
for moderate recombination, given instead high enough mutation rates.

\section{Approximate inference in QLE}\label{app:Inference_in_QLE}
QLE gives quantitative relations between evolutionary dynamics
and distributions over genotypes in a population.
In this section we state and briefly describe what these
relations are, and how we use them in this work.

We are only here concerned with the most important relation,
which
connects the parameters $J_{ij}$ in \eqref{eq:Ising}
to parameters describing
how fitness and recombination shape the population.
Fitness is here taken as a propensity of a genotype
to propagate in the next generation.
It is parametrized as
\begin{equation}
\label{eq:fitness}
F(\mathbf{\sigma})= \sum_i f_i s_i + \sum_{ij} f_{ij}s_is_j
\end{equation}
where the $f_i$ and $f_{ij}$
are respectively the additive and the epistatic components
of fitness and the genome is assumed bi-allelic as
in \eqref{eq:Ising}; for the extension to multi-allelic sites
see \cite{Gao2019}.
The meaning of \eqref{eq:fitness} is that individuals with
higher total fitness (higher $F(\mathbf{\sigma})$) will have higher
expected number of offspring.
Recombination is described by an overall rate $r$
and a locus-pair-dependent quantity $c_{ij}$
which is the probability that alleles at loci $i$ and $j$
are inherited from different parents if a recombination event
has taken place. In many cases it is reasonable to
assume that this quantity is small (close to zero)
when $i$ and $j$ lie sufficiently closely on the genome,
and approximate $\frac{1}{2}$ if they are distant.

The relation first found by Kimura for a two-locus problem is
\begin{equation}\label{eq:Kimura}
J_{ij}= \frac{f_{ij}}{r\cdot c_{ij}}
\end{equation}
In a world abundant in sequenced genomes the left hand side of
above can be inferred from samples. It is thus what can be
considered known, while what causes it, $f_{ij}$ and
the influence of recombination, are quantities not known directly.
It is therefore useful to re-state \eqref{eq:Kimura}
as
\begin{equation}\label{eq:Kimura-inverse}
f^*_{ij}= J^*_{ij}\cdot r c_{ij}
\end{equation}
where the star indicates that these quantities are inferred from data.

Finally, for sites far enough apart on the genome the above
can be summarized that epistatic term in fitness
($f_{ij}$) is proportional to pair-wise term in the
distribution ($J_{ij}$). A ranking of pairs in descending order
of $J_{ij}$ is hence also a ranking of pairs in descending order
of epistatic fitness, if closely spaced pairs are excluded.
This is hence the theoretical basis of the analysis in the main
body of the paper, where we rank pairs as to the values of
$J^*_{ij}$, as inferred from samples.
The extension of the above to Potts genomes
(multi-allelic states) is given in
\cite{Gao2019} and~\cite{Zeng2020}.

\section{Methods of DCA}\label{app:DCA_methods}
Determining the coefficients in a distribution
of the type \eqref{eq:Ising} from data has been
called a problem of \textit{inference in exponential families} in
statistics~\cite{WainwrightJordan2008},
an \textit{inverse Ising problem}
in statistical physics~\cite{Roudi-2009b,Aurell-2012a,Nguyen2017}
and \textit{direct coupling analysis} (DCA)
in computational biology~\cite{MorcosE1293,Cocco2018}.
The benchmark method for such problems is
\textit{maximum likelihood} (ML). This is built from the assumption
that $N$ observed genotypes
are independent draws from a distribution
of type \eqref{eq:Ising}.
The logarithm of the probability of the
joint distribution (log-likelihood) is then
\begin{widetext}
\begin{equation}
    {\cal L}\left(\{h_i\},\{J_{ij}\}\right) = \sum_i h_i \left(\frac{1}{N}\sum_s s_i^{s}\right)
+ \sum_{ij} J_{ij} \left(\frac{1}{N}\sum_s s_i^{s}s_j^{s}\right)
-\log Z
\end{equation}\label{eq:L}
\end{widetext}

The averages on the right hand side multiplying
$h_i$ and  $J_{ij}$ are \textit{empirical averages}.
Since these are the only properties of the samples
which enter into the log-likelihood they are
\textit{sufficient statistics} for ML inference.
The partition function ($Z$) is computationally
difficult to evaluate, and a large number of
other inference procedures
to circumvent this problem
are therefore
widely used, reviewed recently in~\cite{Cocco2018,Nguyen2017}.

One family of widely used approaches is variational methods.
This is based on minimizing the distance between the
empirical distribution and a suitable trial distribution,
in practice mostly factorized distribution~\cite{WainwrightJordan2008},
although later distribution of the type \eqref{eq:Ising} have also been used~\cite{Nguyen2017} (and references therein).
The most widely used variational inference
is \textit{naive mean-field} which is equivalent to treating
the Ising (or Potts) distribution over
discrete variables as if it were a Gaussian distribution
over continuous variables.
While naive mean-field inference was the basis for
important advances in protein structure prediction~\cite{MorcosE1293,Soeding-2017a},
it has more recently been overtaken by pseudo-likelihood
maximization (PLM)~\cite{Besag-1975a,Ravikumar-2010a,Aurell-2012a,Nguyen2017,Cocco2018}. PLM is the DCA method used in this work.
The principles are described in Materials \& Methods section of the
main body of the paper, text around Eqs.~(5) and~(6). The implementation of PLM
which we have used (also called PLM),
due to Chen-Yi Gao~\cite{GaoZhouAurell2018}
is available on~\cite{Gao-github}.

\section{Phylogenetic randomization of DCA: principles}\label{app:randomization}
This section describes the two means of
phylogenetic randomization used in the main paper.
A full-length technical description of both methods is
available (from co-author ERH) at~\cite{RodriguezWeigt-2020}.

\textbf{Profile randomization:} This approach is built on the
principle of randomizing the input multi-sequence alignment (MSA) by conserving the single-columns statistics $w_i(\sigma)$, for  all sites $i=1,...,L$ and all nucleotides or gaps $\sigma\in\left\lbrace -,N,A,C,G,T\right\rbrace $. This is done by independent columns shuffling which destroys all kinds of correlations (both coevolutionary ones and phylogenetic ones) present in the alignment, only the residue conservation patterns of the original MSA are preserved. The randomized sequences become an independently and identically distributed sample from the profile model
$$P(\sigma_1,...,\sigma_L)=\prod_{i=1}^{L}w_i(\sigma_i).$$
Profile randomization hence serves to verify that the effects found
are not due to random sampling.
An alternative way to reach the same goal, and high-lighted in the
main body of the paper, is to repeat the analysis with
successively larger collections of genomes obtained from
using successively later cut-off dates.

\textbf{Phylogeny randomization}:
This more advanced approach is built on the
principle of randomizing the input MSA preserving both single columns statistics and pairwise Hamming distances between sequences representing the genotypes.
This method was first presented in detail in~\cite{RodriguezWeigt-2020}.
Computational methods to infer phylogeny which
rely on such pair-wise Hamming distances would
by design be insensitive to phylogeny randomization,
i.e. the give the same result as using input MSA.
Phylogeny randomization hence serves to distinguish
co-variation between loci due to phylogeny (inheritance) from
co-variation due to co-evolution (epistasis).

The method is initialized with an alignment resulting from the ``profile'' randomization, eliminating all preexisting correlations, to then start a simulated annealing based method to construct a sequence alignment with inter-sequences distances of the original MSA as target.  Single variables  $\sigma^{(i)}_k$ are  permuted in the following way: at each move t, a column $k$ and two rows $m$ and $n$ are chosen at random, and an attempt to exchange $\sigma^{(m)}_k$ and $\sigma^{(n)}_k$ is made. The probability of the exchange to take place is the Metropolis-Hastings acceptance probability:
\begin{widetext}
\begin{equation}
    P_{acc}(t,\beta)=\min\left(1, \exp\left( \beta \left( ||\bm{h}^{t}-\bm{h}_{target}||-||\bm{h}^{t-1}-\bm{h}_{target}||\right) \right)  \right)
\end{equation}
\end{widetext}
where $\bm{h}^{t-1}$ and $\bm{h}^{t}$ are the Hamming distance matrices of the current alignment before and after the exchange, $\bm{h}_{target}$ the Hamming distance matrix corresponding to the original  alignment, $|| \dots ||$ stands for the Frobenius matrix norm, and $\beta$ is an inverse temperature parameter. Thus, a move is more likely to be accepted if it makes the Hamming distance  matrix of the alignment closer to that of the target. Parameter $\beta$ is initialized at a low value and then slowly increased as more moves are made,
When $\beta$ goes to infinity and annealing has proceeded slowly
enough $\bm{h}\rightarrow  \bm{h}_{target}$.

This procedure never changes the single site statistics of the target alignment, since exchanges are made inside one column. On other hand all the epistatic correlations are destroyed. If we assume that the target Hamming distance matrix is a measure of the phylogenetic information in the original MSA, as is done in several tree building methods (distance-based methods)
\cite{Felsenstein2004,Saitou1987}, then we expect that resulting alignment present this hierarchical signal.

\section{Phylogenetic randomization of DCA: results and visualization}\label{app:phylogenetic_results_visulization}
Starting from the filtered MSA for the 2020-08-08 data-set,
we apply
the two randomization strategies described above.
For profile randomization 50 samples are generated and
the PLM procedure is implemented to infer the epistasis between pairwise loci.

\begin{figure*}[htpb]
\centering
\includegraphics[width=0.85\textwidth]{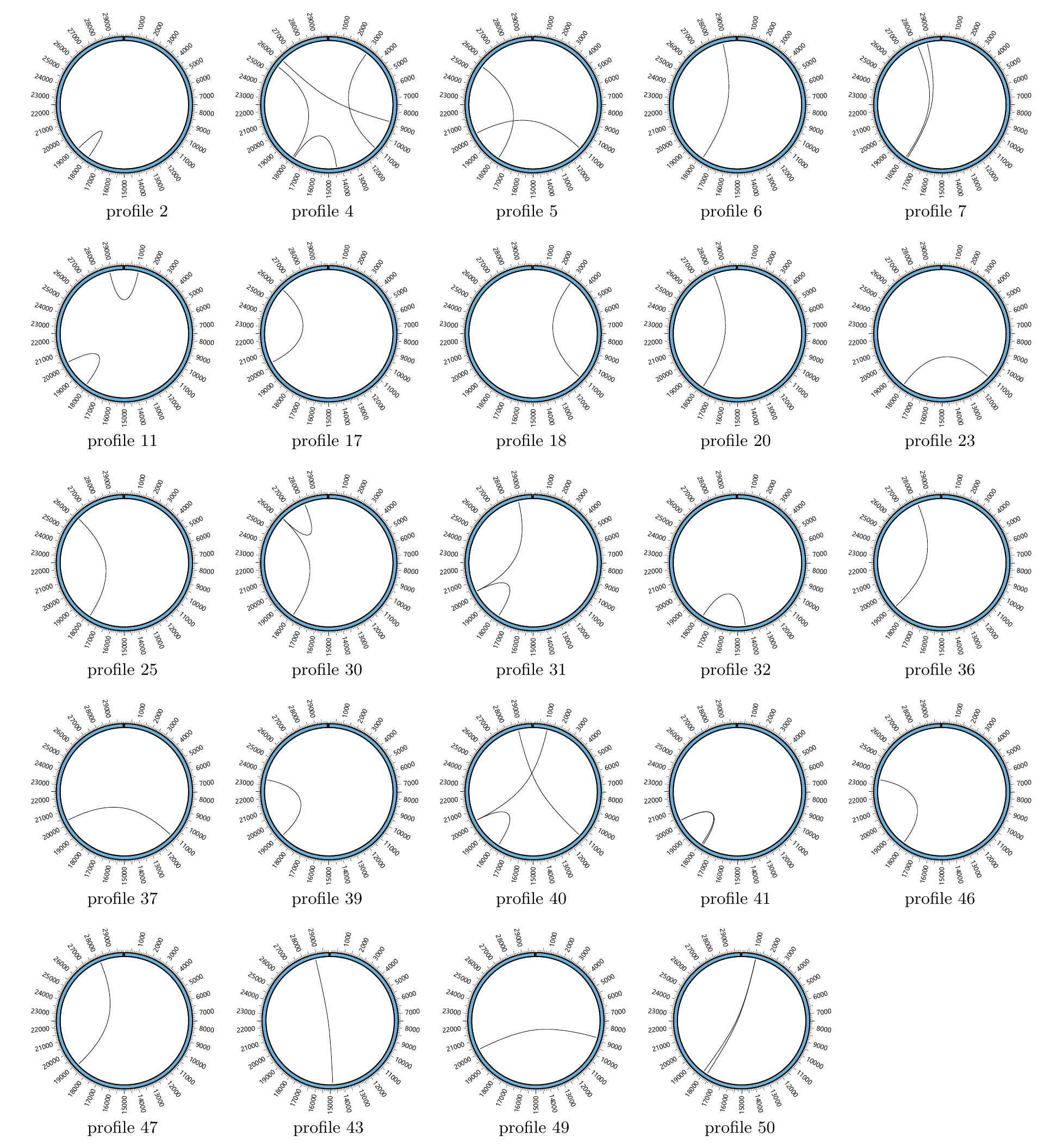}
\caption{Links within top-2000 PLM scores of samples with profile randomization. The links located in coding regions and terminals have mutations excluding gaps and 'N's are retained
in these circos plots. There are 24 out of 50 samples contain links. However, all of these are ranked low starting from the original MSA,
and none of them show up in Table~\ref{tab:tab1} in the main body of the paper.}
\label{figs:profile_samples}
\end{figure*}

As stated in main body of the paper,
for further analysis we only retained
pairs where both the inferred loci are located in the coding region and the first and second prevalent nucleotides of these loci in the analyzed data-set are entries in \{A,C,G,T\} .
We observed no such pairs in the top-200 PLM scores, for any profile randomized samples.
Therefore, the examined ranks of PLM scores are extended to top-2000s. In Fig. \ref{figs:profile_samples}, the inferred top-2000 epistasis for each sample are shown. There are 24 out of 50 samples which contain interactions
found by PLM. However, \textbf{as they all have low rank}, none of these links show up in Table~\ref{tab:tab1} in the main body of the paper.

\begin{figure*}[htpb]
\centering
\includegraphics[width=0.85\textwidth]{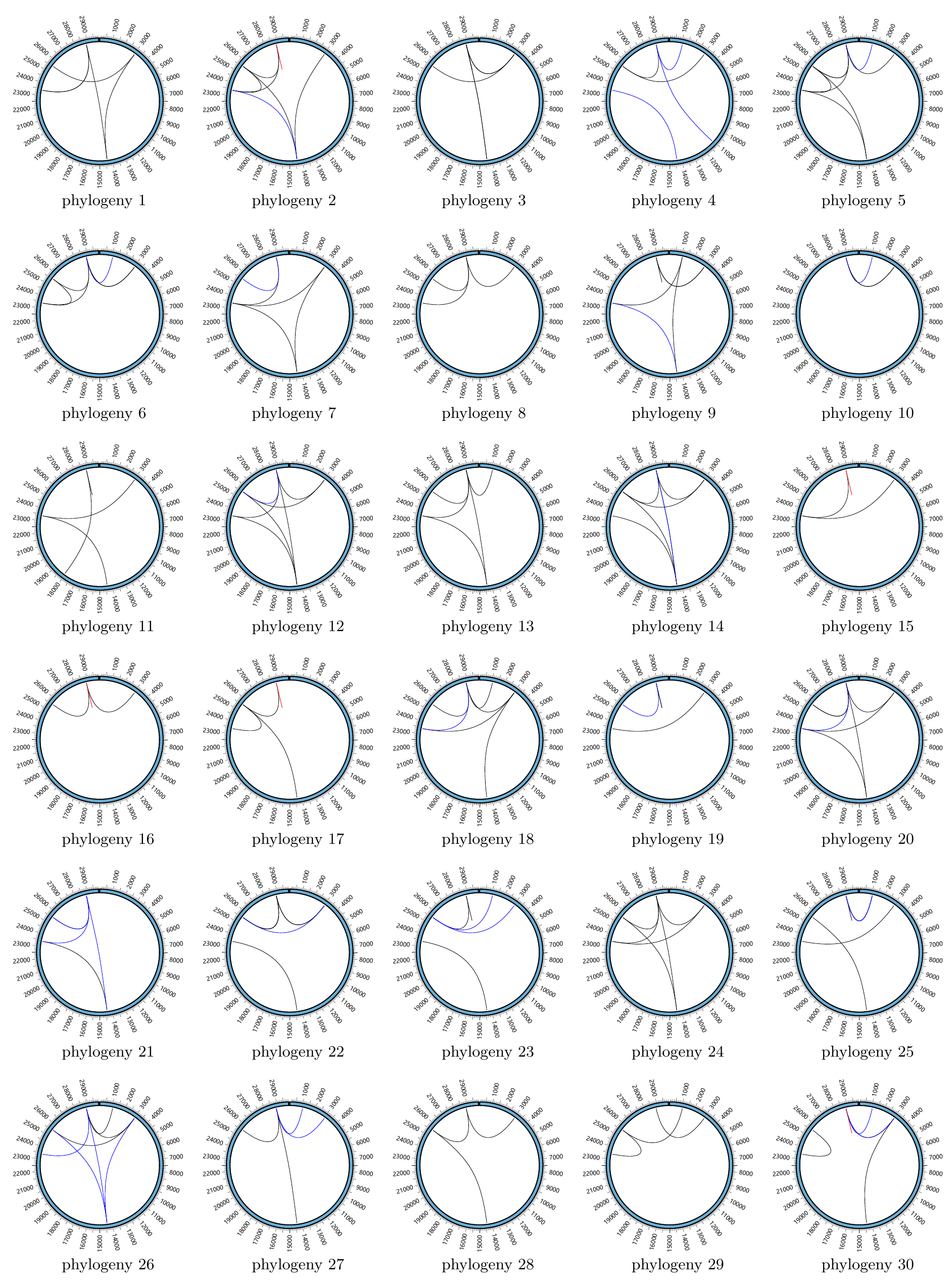}
\caption{Links within top-200 PLM scores of the 1st to the 30th samples with phylogenetic randomization. The links located in coding regions are retained
in these circos plots.
Each phylogenetic sample contains interactions effects with a rank cut-off threshold of 200. Blue shows links with lengths $d_{ij}$ longer than 3 nucleotide. Red shows links with $d_{ij}\le 3 nt$  for links with rank in top-50s. Black shows links with $d_{ij}\le 3 nt$  for links with rank in top 51-200.}\label{figs:phylogenetic_samples}
\end{figure*}

\begin{figure*}[htpb]
\centering
\includegraphics[width=0.85\textwidth]{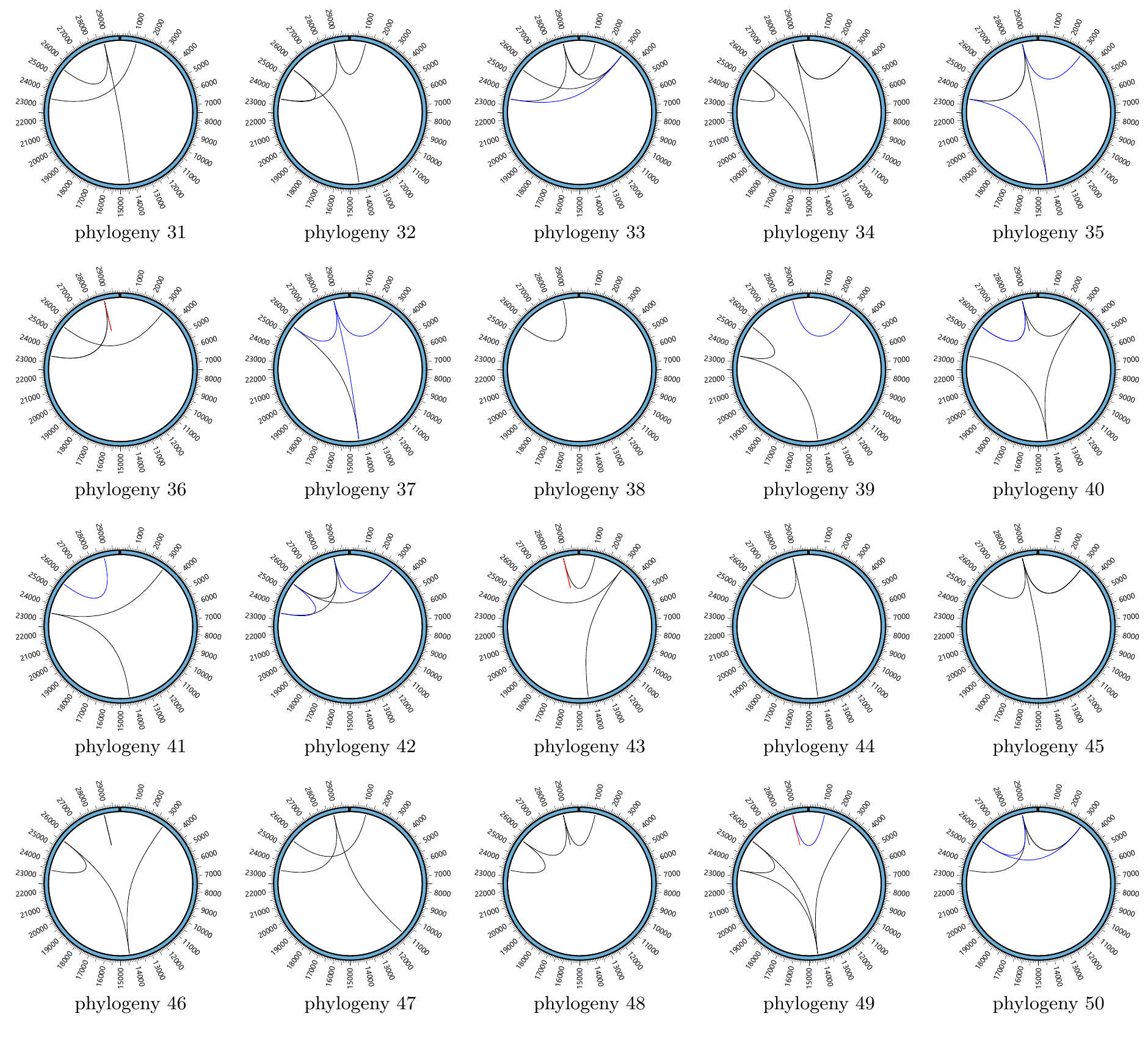}
\caption{Links within top-200 PLM scores of the
31st to 50th samples with phylogenetic randomization.}
\label{figs:phylogenetic_samples_2}
\end{figure*}

For phylogeny randomization we also generated 50 random samples. The inferred epistasis for these samples are shown in Fig.~\ref{figs:phylogenetic_samples} and continued in Fig.~\ref{figs:phylogenetic_samples_2}. Unlike profile strategy, a subset of the
putative epistatic links of Table \ref{tab:tab1} in the main body of the paper
also show up after phylogeny randomization.
This subset of predictions is listed in
Table \ref{tab:tab2} in the main body of the paper,
and has been eliminated in the list of
retained predictions in Table \ref{tab:tab3}.


\section{Robustness as to thresholds}\label{app:robustness_threshold}
\begin{figure}[htpb]
\centering
\includegraphics[width=0.4\textwidth]{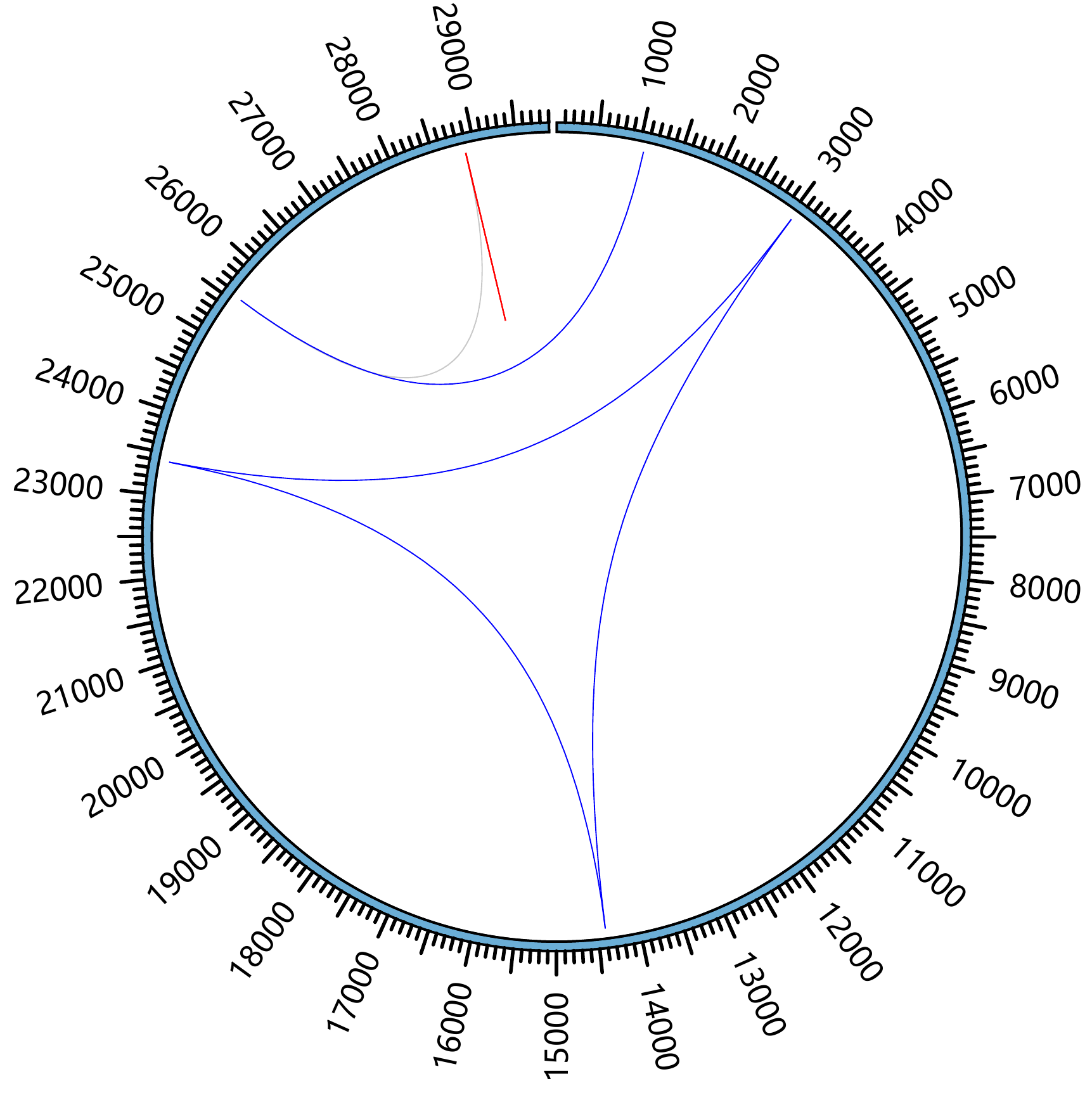}\\
\includegraphics[width=0.4\textwidth]{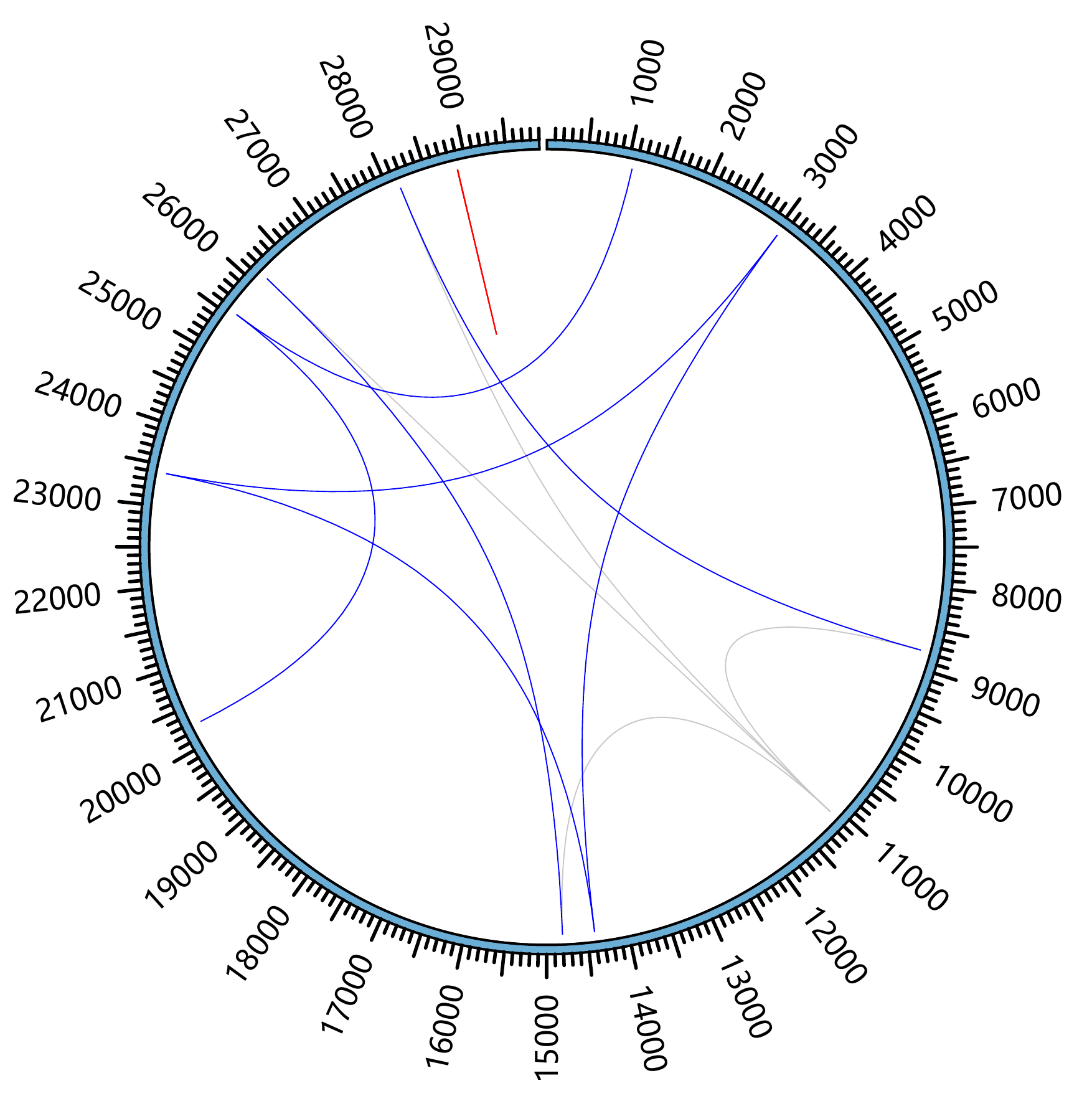}
\caption{Top-200 significant pairwise epistasis from the 2020-08-08 data-set. Upper: the filtering threshold is 90\% (for a given locus, if the percentage of a certain main nucleotide (A, C, G, T, -, N) exceeds the threshold, this locus will be deleted.); Lower: the threshold value is 95\%. Only epistatic links with both terminals located in the coding region and no gaps and 'N's included in mutations are presented here. Colored lines indicate for top 50s, grey for links with ranks in top 50-200s. Red lines show short-distance links ($\le$ 3 bp) blue lines show links of longer distance.}
\label{thresholds}
\end{figure}

In the analysis in main body of the paper, the following criteria are used for pre-filtering of the MSA. First, if in one locus the same nucleotide is found in a percentage $ P_{major}\ge 96.5\%$ of the samples, or if the sum of the percentages of A, C, G and T at this position is $P_{sum}\le 20\%$, then this locus will be deleted. The inferred epistasis effects are found to be sensitive to $P_{major}$ at a certain locus but not to $P_{sum}$.
We chose the criterion $P_{major} \ge 96.5\%$ to keep
consistency with earlier obtained results from the dataset of 2020-05-02 (presented below).
For that earlier (smaller) dataset, the same results are obtained
also with less permissive thresholds.
To check the influence of $P_{major}$, we performed PLM inference over the 2020-08-08 dataset with $P_{major}=90\%$ and $P_{major}=95\%$ respectively. The inferred epistasis for these two $P_{major}$s are shown in the left and right panel of Fig.~\ref{thresholds} respectively. By increasing the threshold, more epistasis effects are inferred.
In summary, most of the links presented with lower $P_{major}$  remain in the results with higher thresholds.

For the visualization of epistasis effects, we focus on links with both terminals located in coding regions and mutation types are gaps and 'N's excluded. As shown in Fig.~\ref{thresholds}, different colors mean links with different ranks. Red (links with terminals located close to each other) and blue (the terminals are far away from each other) are for the epistasis links with ranks less than 50 while grey for links with ranks within 51 to 200. Both panels clearly show that with PLM method, most significant links are in top-50s. This indicates that the choice of visualization threshold are reasonable in the presentation of epistasis effects.

\section{Different quantifications of correlations}\label{app:quantification_correlation}

The correlation (co-variance) of two
Boolean variables is given by one real number.
If the Boolean variables are
represented as spin variables $s_i$ and $s_j$
taking values $1$ and $-1$, this number is
the physical correlation
$\chi_{ij}=\langle s_i s_j \rangle - \chi_i \chi_j$,
where $\chi_i =\langle s_i\rangle$ and $\chi_j = \langle s_j \rangle$.
Since every joint distribution on two Boolean variables
can be written
$p(s_i,s_j) = \frac{1}{4}\left(1+\chi_i s_i +\chi_j s_j + (\chi_i\chi_j+\chi_{ij})s_is_j\right)$
with marginals
$p(s_i) = \frac{1}{2}\left(1+\chi_i s_i \right)$
and
$p(s_j) = \frac{1}{2}\left(1+\chi_j s_j \right)$,
the contrast is
\begin{eqnarray}
\label{eq:contrast}
    c(s_i,s_j)&=&\frac{p(s_i,s_j)}{p(s_i)p(s_j)}\nonumber\\
    &=& \frac{1+\chi_i s_i +\chi_j s_j + (\chi_i\chi_j+\chi_{ij})s_is_j}{1+\chi_i s_i +\chi_j s_j + \chi_i\chi_j s_is_j}~.
\end{eqnarray}
The mutual information (MI),
recently used in genome-scale epistasis
analysis in~\cite{Pensar2019},
is the expected log-contrast
\begin{eqnarray}
     \label{eq:MI}
    I_{ij} &=& \sum_{s_i,s_j} p(s_i,s_j) \log c(s_i,s_j) \nonumber\\
&=& \frac{1}{4}\left( (1+\chi_i + \chi_j + a_{ij})
    \log \frac{1+\chi_i +\chi_j + a_{ij}}{1+\chi_i + \chi_j + \chi_{i}\chi_j}\right.
    \nonumber \\
&&  + (1-\chi_i + \chi_j -a_{ij})
\log \frac{1-\chi_i +\chi_j -a_{ij}}{1-\chi_i + \chi_j -\chi_i\chi_j}\nonumber \\
&& + (1+\chi_i -\chi_j -a_{ij})
  \log \frac{1+\chi_i -\chi_j - a_{ij}}{1+\chi_i -\chi_j -\chi_{i}\chi_j}
    \nonumber \\
&& + \left.(1-\chi_i - \chi_j +a_{ij})
     \log \frac{1-\chi_i -\chi_j +a_{ij}}{1-\chi_i - \chi_j +\chi_{i}\chi_j}  \right)~.\nonumber\\
\end{eqnarray}
with $a_{ij}=\chi_i\chi_j+\chi_{ij}$.\\
For given magnetizations ($\chi_i$ and $\chi_j$),
correlations and
mutual information
of Boolean variables
are hence in one-to-one correspondence.
In particular, zero correlation implies zero MI.
For categorical data (more than two states per variable),
correlation is conveniently defined as a matrix
\begin{equation}
f_{ij}(a,b)=\langle \mathbf{1}_{s_i,a} \mathbf{1}_{s_j,b} \rangle - f_i(a) f_j(b)
\end{equation}
where $f_{i}(a)=\langle \mathbf{1}_{s_i,a} \rangle$ and
$f_{j}(b)=\langle \mathbf{1}_{s_j,b} \rangle$.
Mutual information is defined in the same way as
in \eqref{eq:MI}, except that the sums go over the
ranges of indices $a$ and $b$.
Frobenius norm (or score) of a correlation matrix is defined as
\begin{equation}
s_{ij} = \sqrt{\sum_{a,b} f_{ij}^2(a,b)}
\end{equation}

\begin{figure}[htpb]
    \centering  \includegraphics[width=0.45\textwidth]{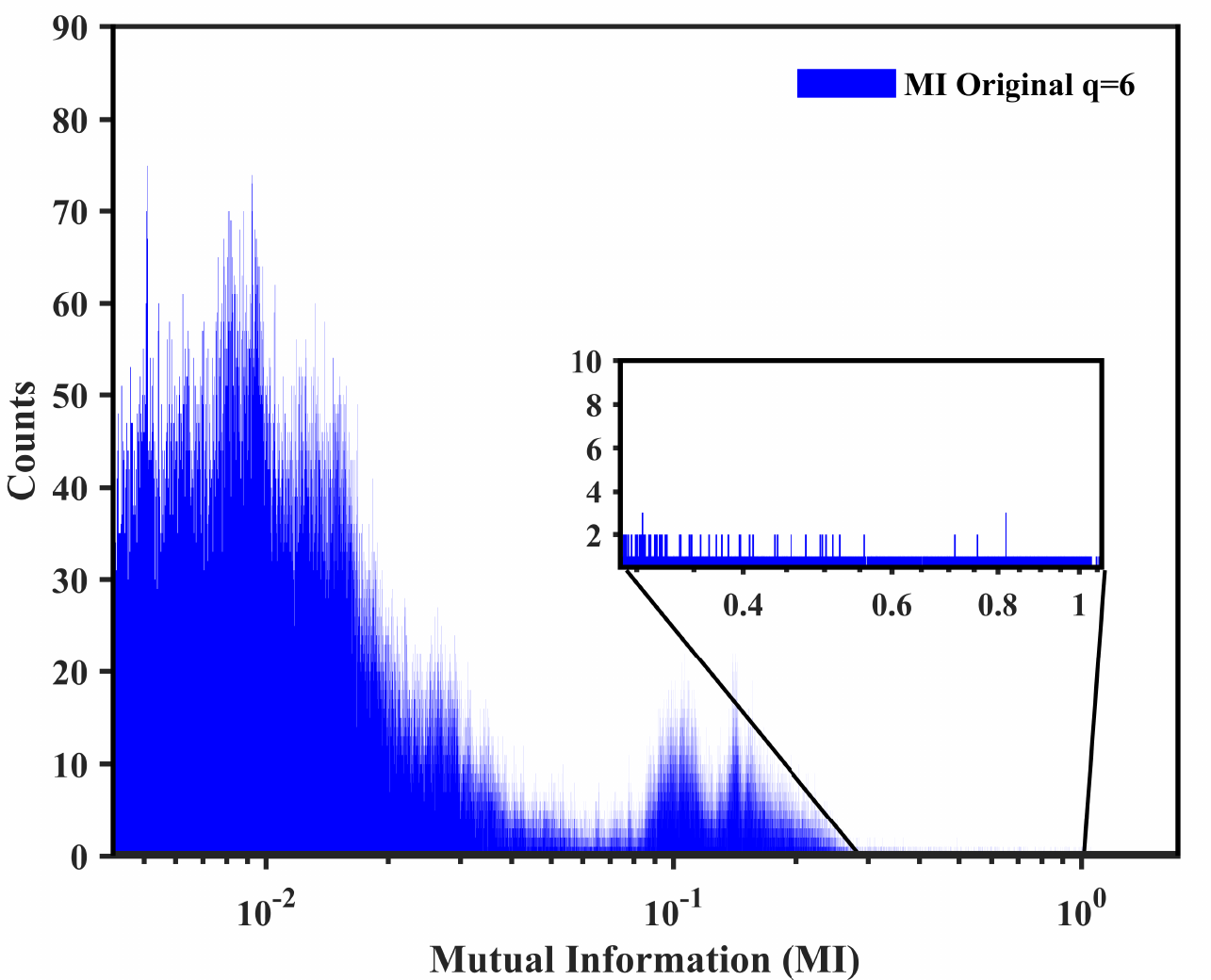}
    {\color{red} \put(-111,90){\huge{$\downarrow$}}}
    \caption{Distribution of Mutual information (MI) between pairwise loci for the original data-set with cut-off date 2020-08-08. The score pointed by the red arrow corresponds to the link with most significant score (1059 to 25563) by PLM analysis.}
    \label{fig:MI_hist}
\end{figure}

Mutual information and Frobenius norm
of correlations of categorical variables
are not generally related. It could therefore be
the case that the information Fig. (3)
and Table~\ref{tab:tab4} in main body of the paper
would be different if the assessment
would be done for mutual information.
Fig.~\ref{fig:MI_hist} and Table~\ref{tab:MI_top10} show that this is substantially not
the case.

\begin{table*}[htpb]
\caption{Top-10 significant links found by mutual information (MI) analysis in the coding region for the dataset with cut-off date 2020-08-08.}
\centering
\begin{ruledtabular}
\begin{tabular}{lllc}
\textrm{Rank \footnote{ Rank for top-10 coding links highlighted by mutual information (MI) analysis. Epistatic links with terminals having synonymous mutations or located in the non-coding regions are omitted.}} & Locus 1 -  & Locus 2 - & Mutual  \\
~ &  protein &  protein  & Information\\
\hline
782 &  28882-N   &   28883-N  &	0.5752\\
792 &  28881-N   &   28882-N  & 0.5698\\
794 &  28881-N   &   28883-N  &	0.5689\\
824 &  3037-nsp3  &  23403-S  & 0.5552 \\
836 &  3037-nsp3  &  14408-nsp12  & 0.5478 \\
840 &  14408-nsp12  & 23403-S & 0.5455\\
1643 & 1059-nsp2 & 25563-ORF3a & 0.3261\\
1775 & 8782-nsp4 & 28144-ORF8 & 0.3118\\
18484 & 17858-nsp13 & 18060-nsp14 & 0.1705\\
19364 & 17747-nsp13 & 17858-nsp13  & 0.1679\\
\end{tabular}\\
\end{ruledtabular}
\label{tab:MI_top10}
\end{table*}


Another way to quantify
interdependence between two
random variables
is the $p$-value of Fisher's exact test~\cite{Fisher-exact-test},
recently used in the present context in~\cite{Cui2019}.
If in $n$ samples in total
outcome $ab$ is found $n_{ab}$ time then
the $p$-value of Fisher's exact test
is the probability that these outcomes
would have been observed in independent
draws of independently distributed variables.
For Boolean variables that has
the exact expression

\begin{widetext}
\begin{equation}
p_{ij}(n, \{n_{ab}\})= \frac{(n_{11}+n_{1-1})!\cdot (n_{-11}+n_{-1-1})!\cdot (n_{11}+n_{-11})!\cdot
              (n_{1-1}+n_{-1-1)})!}
{n_{11}!\cdot n_{-11}!\cdot n_{1-1}!\cdot n_{-1-1}!\cdot n!}
\end{equation}
\end{widetext}
In the limit of many samples $n_{ab}=n\cdot p_{ij}(a,b)$
almost surely, and by Stirling's formula
\begin{widetext}
\begin{eqnarray}
\log p_{ij}(n) &\approx& n\big(
     (p_{ij}(1,1)+p_{ij}(1,-1))\log (p_{ij}(1,1)+p_{ij}(1,-1))
    + (p_{ij}(-1,1)+p_{ij}(-1,-1))\log (p_{ij}(-1,1)+p_{ij}(-1,-1)) \nonumber \\
& +& (p_{ij}(1,1)+p_{ij}(-1,1))\log (p_{ij}(1,1)+p_{ij}(-1,1))
 + (p_{ij}(1,-1)+p_{ij}(-1,-1))\log (p_{ij}(1,-1)+f_{ij}(-1,-1)) \nonumber \\
& -& p_{ij}(1,1)\log p_{ij}(1,1) - p_{ij}(1,-1) \log p_{ij}(1,-1)
         - p_{ij}(-1,1)\log p_{ij}(-1,1) - p_{ij}(-1,-1) \log p_{ij}(-1,-1) \big), \nonumber \\
\end{eqnarray}
\end{widetext}
which is mutual information, up to a factor $-n$.
In a more general setting, this
result follows from Sanov's lemma
in Information Theory.
For the data we consider we are always in the range of very
large $n$, and Fisher's exact
test therefore does not give additional information.

\section{On putative couplings between nsp7 (11843..12091) and nsp8 (12092..12685)}\label{app:nsp7_nsp8}
As described above, the set of loci remaining after filtering depends on the percentage of the a major nucleotide $P_{major}$
used as threshold. As the two viral genes nsp7 and nsp8 are known to interact \cite{Peng2020}，
we looked for epistatic interactions between loci in these two genes.
None appear using the threshold $P_{major}=96.5\%$  employed in the main body of the paper.
All loci located in nsp7 and nsp8 are deleted during filtering procedure using this threshold.
To nevertheless consider possible epistasis effects within these two regions, we increased the value of $P_{major}$. As $P_{major}$ grows more loci remain
after filtering, in all regions, which increases the computational burden. To
mitigate this effect, we further filtered the MSA matrix
by considering the mutation type of each remaining loci.
In the following we have filtered out loci where one of the
dominant mutation type are gaps ('-') or unknown nucleotide ('N').

With these alternative filtering criteria, we tested values of $P_{major}$
from 97\% to 99.9\% and found that the loci in nsp7 and nsp8
show up when $P_{major} > 99.7\%$. With $P_{major} = 99,8\%$,
there remains 262 loci and 5,746 unique sequences,
 from which one locus at position 11916 in nsp7 and two loci
 at positions 12199 and 12478 in nsp8 are identified.
 With $P_{major} = 99,9\%$, there remains 580 loci and 9525 unique sequences,
from which two loci at positions 11916 and 12025 in nsp7
are identified and seven closely positioned loci (12199, 12400, 12478, 12503, 12513, 12534, 12550) in nsp8.

With $P_{major} =99.8\%$, there are 34,191 inferred pairwise epistatic links in total.  With $P_{major} =99.9\%$, the number of inferred links is 167,910. The epistasis analysis results provided by PLM procedure are summarized in Table~\ref{tab:tab_nsp7_8}. Some significant links with high ranks are included
for comparison. Both ranks and PLM scores show that the epistasis effects between nsp7 and nsp8 are much weaker than the significant ones.
The ranks for the listed links in Table~\ref{tab:tab_nsp7_8} are obtained without considering gaps '-' and not recognized notation 'N'. However, they fit well with the results that including the effects of gaps and 'N's, at least for the links with the rank of top-50s.
In summary, links between loci in genes nsp7 and nsp8
only appear using much more permissive filtering criteria
and with much lower rank than the top-200 listed
in Table~\ref{tab:tab1} in the main body of the paper.

\begin{table*}[!ht]
\centering
\caption{Links between loci in genes nsp7 and nsp8
obtained after the modified filtering procedure applied to the 2020-08-08 data set.
Loci without gaps '-', without not recognized 'N' and satisfying
permissive threshold criterion $P_{major} = 99.8\%$ and $P_{major} = 99.9\%$  have been retained and reported here.}
\begin{ruledtabular}
\begin{tabular}{llllllll}
Locus 1 & mutation  & Locus 2  & mutation & PLM score  & Rank & PLM score  & Rank  \\
-protein & -type & -protein & -type & \multicolumn{2}{c}{$P_{major} = 99.8\%$ } & \multicolumn{2}{c}{$P_{major} = 99.9\%$}\\
\hline
1059-nsp2 & C$|$T-non.  & 25563-ORF3a & G$|$T-non. & 1.9526 & 3 & 2.0365 & 1 \\

8782-nsp4 & C$|$T-syn. & 28144-ORF8 & T$|$C-non. & 1.4479 & 6 & 1.9739 & 6 \\

17747-nsp13 & C$|$T-non. & 17858-nsp13 & A$|$G-non. & 0.8553 & 27 & 1.0522 & 22 \\
17858-nsp13 & T$|$C-non. & 18060-nsp14 & C$|$T-syn. & 0.8624 & 26 & 0.9787 & 27\\
17747-nsp13 & C$|$T-non. & 18060-nsp14 & C$|$T-syn. & 0.7780 & 36 & 0.8529 & 39 \\
11083-nsp6 & G$|$T-non. & 14805-nsp12 & C$|$T-syn. & 0.5040 & 134 & 0.7109 & 58 \\
\hline
11916-nsp7 & C$|$T-non.  & 12199-nsp8 & A$|$G-syn.  & 0.0259  &  19781  & 0.0113  &  86504 \\

11916-nsp7 & C$|$T-non. & 12478-nsp8 & G$|$A-non. & 0.0211  & 25861  & 0.0097  & 109600 \\
\hline
12025-nsp7 & C$|$T-syn.  & 12550-nsp8 & G$|$A-syn.  &  ~  & ~    & 0.0325  &  20933 \\

12025-nsp7 & C$|$T-syn. & 12199-nsp8 & A$|$G-syn. & ~  & ~  & 0.0264  & 29496 \\

12025-nsp7 & C$|$T-syn. & 12534-nsp8 & C$|$T-non. & ~  & ~  & 0.0237  & 35797 \\

12025-nsp7 & C$|$T-syn. & 12478-nsp8 & G$|$A-non. & ~  & ~  & 0.0236  & 35936 \\

12025-nsp7 & C$|$T-syn. & 12400-nsp8 & C$|$T-syn. & ~  & ~  & 0.0202  & 48464 \\

11916-nsp7 & C$|$T-non. & 12400-nsp8 & C$|$T-syn. & ~  & ~  & 0.0119  & 83501 \\

11916-nsp7 & C$|$T-non. & 12513-nsp8 & C$|$T-syn. & ~  & ~  & 0.0117  & 84562 \\

11916-nsp7 & C$|$T-non. & 12534-nsp8 & C$|$T-non. & ~  & ~  & 0.0110  & 89102 \\
11916-nsp7 & C$|$T-non. & 12503-nsp8 & T$|$C-non. & ~  & ~  & 0.0106  & 92834 \\

12025-nsp7 & C$|$T-syn. & 12513-nsp8 & C$|$T-non. & ~  & ~  & 0.0099  & 103746 \\

11916-nsp7 & C$|$T-non. & 12550-nsp8 & G$|$A-syn. & ~  & ~  & 0.0098  & 105534 \\

12025-nsp7 & C$|$T-syn. & 12503-nsp8 & T$|$C-non. & ~  & ~  & 0.0098  & 106633 \\

\end{tabular}\\
\end{ruledtabular}
\label{tab:tab_nsp7_8}
\end{table*}

\begin{table*}[htpb]
\centering
\caption{Links between loci in genes nsp10 and nsp14
obtained after the modified filtering procedure applied to the 2020-08-08 data set.
Loci without gaps '-', without not recognized 'N' and satisfying
permissive threshold criterion $P_{major} = 99.8\%$ and $P_{major} = 99.9\%$  have been retained and reported here. Only links appears in the filtering by both values of threshold are presented here.}
\begin{ruledtabular}
\begin{tabular}{llllllll}
Locus 1 & mutation  & Locus 2  & mutation & PLM score  & Rank & PLM score  & Rank  \\
-protein & -type & -protein & -type & \multicolumn{2}{c}{$P_{major} = 99.8\%$ } & \multicolumn{2}{c}{$P_{major} = 99.9\%$}\\
\hline
1059-nsp2 & C$|$T-non.  & 25563-ORF3a & G$|$T-non. & 1.9526 & 3 & 2.0365 & 1 \\
8782-nsp4 & C$|$T-syn. & 28144-ORF8 & T$|$C-non. & 1.4479 & 6 & 1.9739 & 6 \\
17747-nsp13 & C$|$T-non. & 17858-nsp13 & A$|$G-non. & 0.8553 & 27 & 1.0522 & 22 \\
17858-nsp13 & T$|$C-non. & 18060-nsp14 & C$|$T-syn. & 0.8624 & 26 & 0.9787 & 27\\
17747-nsp13 & C$|$T-non. & 18060-nsp14 & C$|$T-syn. & 0.7780 & 36 & 0.8529 & 39 \\
11083-nsp6 & G$|$T-non. & 14805-nsp12 & C$|$T-syn. & 0.5040 & 134 & 0.7109 & 58 \\
\hline
13216-nsp10 & T$|$G-non. & 18060-nsp14 & C$|$T-syn. & 0.0407 & 10743 & 0.0201 & 48930 \\
13216-nsp10 & T$|$G-non. & 18788-nsp14 & C$|$T-non. & 0.0258 & 19879 & 0.0115 & 85522 \\
13216-nsp10 & T$|$G-non. & 18877-nsp14 & C$|$T-syn. & 0.0222 & 23695 & 0.0120 & 83143 \\
\end{tabular}\\
\end{ruledtabular}
\label{tab:tab_nsp10_14}
\end{table*}

\section{On putative couplings between nsp10 (13025..13441) and nsp14 (18040..19620)}\label{app:nsp10_nsp14}
As the two viral genes nsp10 and nsp14 are also known to interact \cite{Ma2015,Romano2020}
we looked for epistatic interactions between loci in these two genes,
applying the same procedure as above.
With $P_{major}=99.8\%$ and $P_{major}=99.9\%$, we show three links between
loci located in nsp10 and nsp14 that show up in the filtering MSAs with both values of threshold. The corresponding PLM scores and the rank are given in Table~\ref{tab:tab_nsp10_14}. Similarly to the links
between loci in nsp7 and nsp8, it has very low rank, as well as low PLM score.
In the main body of the paper, where filtering
criterion $P_{major}=96.5\%$ is used, these loci are filtered out
and do not appear.

\section{On putative couplings involving loci in Spike}\label{app:D614G}

\begin{table*}[!ht]
\centering
\caption{Links involving loci in Spike protein
obtained after the filtering procedure described in the main context applied to the 2020-08-08 data set with $P_{major}=96.5\%$.}
\begin{ruledtabular}
\begin{tabular}{llllll}
Locus 1 & mutation  & Locus 2  & mutation & PLM score  & Rank  \\
-protein & -type & -protein & -type & \multicolumn{2}{c}{$P_{major} = 96.5\%$ } \\
\hline
3037-nsp3	& T$|$C-syn.	& 23403-S		& G$|$A-non.	& 1.0114	& 14\\
14408-nsp12	& T$|$C-non.	& 23403-S		& G$|$A-non.	& 0.9917	& 17\\
23403-S		& G$|$A-non.	& 25563-ORF3a	& G$|$T-non.	& 0.3440	& 367\\
11083-nsp6	& G$|$T-non.	& 23403-S		& G$|$A-non.	& 0.3246	& 428\\
23403-S		& G$|$A-non.	& 28144-ORF8	& T$|$C-non.	& 0.3240	& 430\\
8782-nsp4	& C$|$T-syn.	& 23403-S		& G$|$A-non.	& 0.3022	& 522\\
14805-nsp12	& C$|$T-syn.	& 23403-S		& G$|$A-non.	& 0.2787	& 672\\
20268-nsp15	& A$|$G-syn.	& 23403-S		& G$|$A-non.	& 0.2496	& 964\\
23403-S		& G$|$A-non.	& 26144-ORF3a	& G$|$T-non.	& 0.2349	& 1166\\
23403-S		& G$|$A-non.	& 28881-N		& G$|$A-non.	& 0.2108	& 1571\\
23403-S		& G$|$A-non.	& 28882-N		& G$|$A-syn.	& 0.2093	& 1607\\
23403-S		& G$|$A-non.	& 28883-N		& G$|$C-non.	& 0.2083	& 1619\\
1059-nsp2	& C$|$T-non.	& 23403-S		& G$|$A-non.	& 0.2000	& 1798\\
18060-nsp14	& C$|$T-syn.	& 23403-S		& G$|$A-non.	& 0.1603	& 2844\\
17858-nsp13	& A$|$G-non.	& 23403-S		& G$|$A-non.	& 0.1532	& 3077\\
17747-nsp13	& C$|$T-non.	& 23403-S		& G$|$A-non.	& 0.1392	& 3618\\
18877-nsp14	& C$|$T-syn.	& 23403-S		& G$|$A-non.	& 0.1186	& 4523\\
\end{tabular}\\
\end{ruledtabular}
\label{tab:tab_Spike_protein}
\end{table*}
D614G in Spike is a well known mutation \cite{Korber2020} of SARS-CoV-2, and has become the most prevalent form in the global pandemic COVID19. This mutation is at position of 23403 with respect to the reference genomic sequence (Wuhan-Hu-1). Through PLM procedure on the whole genomic sequences, we observed two pairwise couplings 3037-23403 and 14408-23403 in top-50 PLM scores. These two pairwise links however showed fairly often in phylogeny randomization tests,
and have therefore been interpreted as effects of shared inheritance (phylogeny).
In above we show all PLM links involving loci in Spike
up to rank 5000, all of them significantly below top-200.
A notable observation as shown in Table \ref{tab:tab_Spike_protein} is that the locus 23403 appears in all these links.
As the epistatic inference is built on retaining the links
with highest PLM scores that also do not also appear after randomization,
none of the entries in Table~\ref{tab:tab_Spike_protein}
are retained as predicted epistatic interactions in Table~\ref{tab:tab3}
of the main body of the paper.

\section{Potential drugs for proteins in Table \ref{tab:tab3} of the
main body of the paper} \label{app:potential_drags}

There are eight proteins listed in Table \ref{tab:tab3} in main body of
the paper. Except ORF3a, all these viral proteins have potential
drugs listed in \cite{Gordon2020}, sorted by human interactors of these proteins, see Table~\ref{tab:drugs}. As
indicated in the table, several of these drugs are approved, for different purposes listed in~\cite{Gordon2020}, while some are still in pre-clinical or clinical trials.

\begin{table*}[!ht]
\centering
\caption{Potential drugs for interactors of proteins in Table \ref{tab:tab3}, as listed in \cite{Gordon2020}. Approved drugs for any purpose in boldface.}
\begin{ruledtabular}
\begin{tabular}{llll}
Bait & Interactor(s) & Drug & Status\\
\hline
nsp2   & FKBP15       & Rapamycin$^{\rm b}$    & \textbf{Approved} \\
  ~    & EIF4E2/H     & Zotatifin$^{\rm b}$    & Clinical trials \\
ORF3a  & None in \cite{Gordon2020} & - & -\\
nsp4   &  NUPs RAE1   & Selinexor$^{\rm b}$    & \textbf{Approved}\\
ORF8   &  DNMT1       &Azacitidine $^{\rm a}$  & \textbf{Approved}\\
  ~    &  LOX         & CCT 365623 $^{\rm a}$  & Pre-clinical\\
  ~    & FKBP7/10     & Rapamycin$^{\rm b}$    & \textbf{Approved} \\
  ~    & FKBP7, FKBP10& FK-506$^{\rm b}$       & \textbf{Approved} \\
  ~    & PLOD1/2      & Minoxidil$^{\rm b}$    & \textbf{Approved} \\
nsp14  & IMPDH2       & Merimepodib $^{\rm a}$ & Clinical Trial\\
  ~    & GLA          & Migalastat $^{\rm a}$  & \textbf{Approved}\\
  ~    & IMPDH2       & Mycophenolic acid $^{\rm a}$  & \textbf{Approved}\\
  ~    & IMPDH2      & Ribavirin $^{\rm a}$   & \textbf{Approved}\\
  ~ & IMPDH2       & Sanglifehrin A $^{\rm b}$& Pre-clinical\\
nsp12  & RIPK1        & Ponatinib $^{\rm a}$   & \textbf{Approved}\\
nsp13  & PRKACA       & H-89 $^{\rm a}$        & Pre-clinical\\
  ~   & MARK3,TBK1   & ZINC95559591$^{\rm a}$ & Pre-clinical\\
  ~   & CEP250       & WDB002$^{\rm b}$       & Clinical Trial\\
nsp6   & ATP6AP1      & Bafilomycin A1 $^{\rm a}$  &   Pre-clinical \\
  ~   & SIGMAR1      & E-52862$^{\rm a}$      & Clinical trials\\
  ~   & SIGMAR1      & PD-144418$^{\rm a}$    & Pre-clinical \\
  ~   & SIGMAR1      & RS-PPCC$^{\rm a}$      & Pre-clinical \\
  ~   & SIGMAR1      & PB28$^{\rm a}$         & Pre-clinical \\
  ~    & SIGMAR1      & Haloperidol$^{\rm a}$  & \textbf{Approved}\\
  ~   & SLC6A15      & Loratadine$^{\rm a}$   & \textbf{Approved}\\
  ~   & SIGMAR1      & Chloroquine $^{\rm b}$ & \textbf{Approved} \\
\end{tabular}\\
\end{ruledtabular}
\footnotesize{$^{\rm a}$ }Entry taken from \cite{Gordon2020}, Supplementary Table 4, "Literature-derived drugs and reagents that modulate SARS-CoV-2 interactors".\\
\footnotesize{$^{\rm b}$ }Entry taken from \cite{Gordon2020}, Supplementary Table 5, "Expert-identified drugs and reagents that modulate SARS-CoV-2 interactors".
\label{tab:drugs}
\end{table*}

\section{Results from data set until 20200502}\label{app:20200502}
As more data accumulates the predictions obtained
from DCA may change.
In the main body of the paper we
show in Fig.~4 that the leading
predictions are stable using four different
cut-off dates:
(April 1, April 8, May 2, August 8).

Here we show as a further robustness test
the other figures and tables of the main
body of the paper, but for the second
largest data set (cut-off date May 2, 2020).

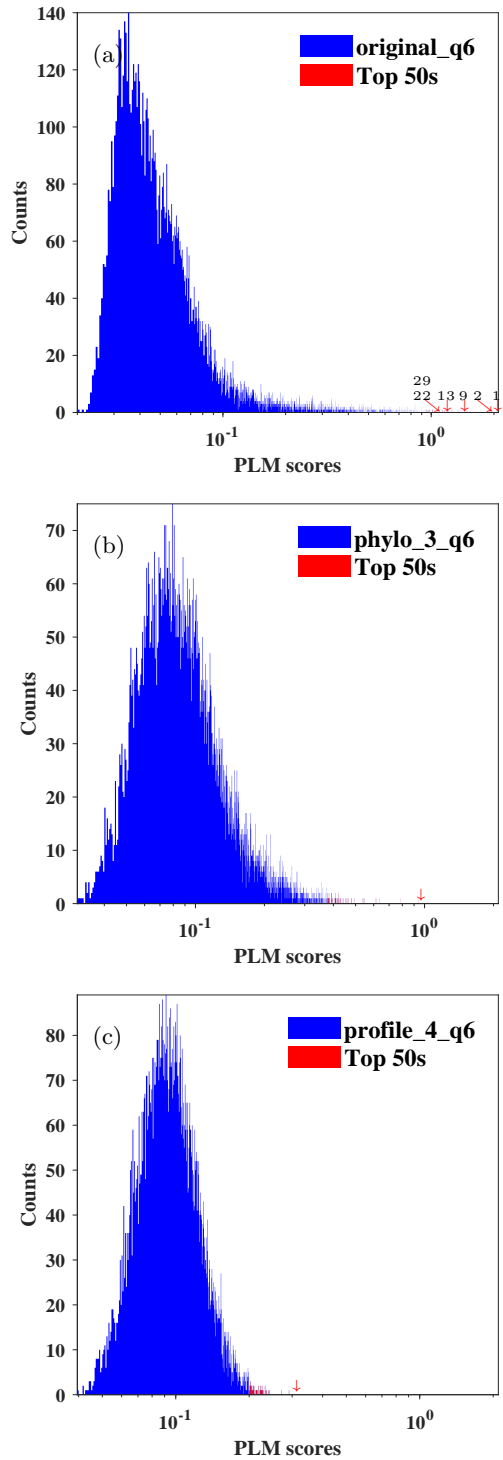
\begin{figure}[htbp]
\centering
\includegraphics[width=0.4\textwidth]{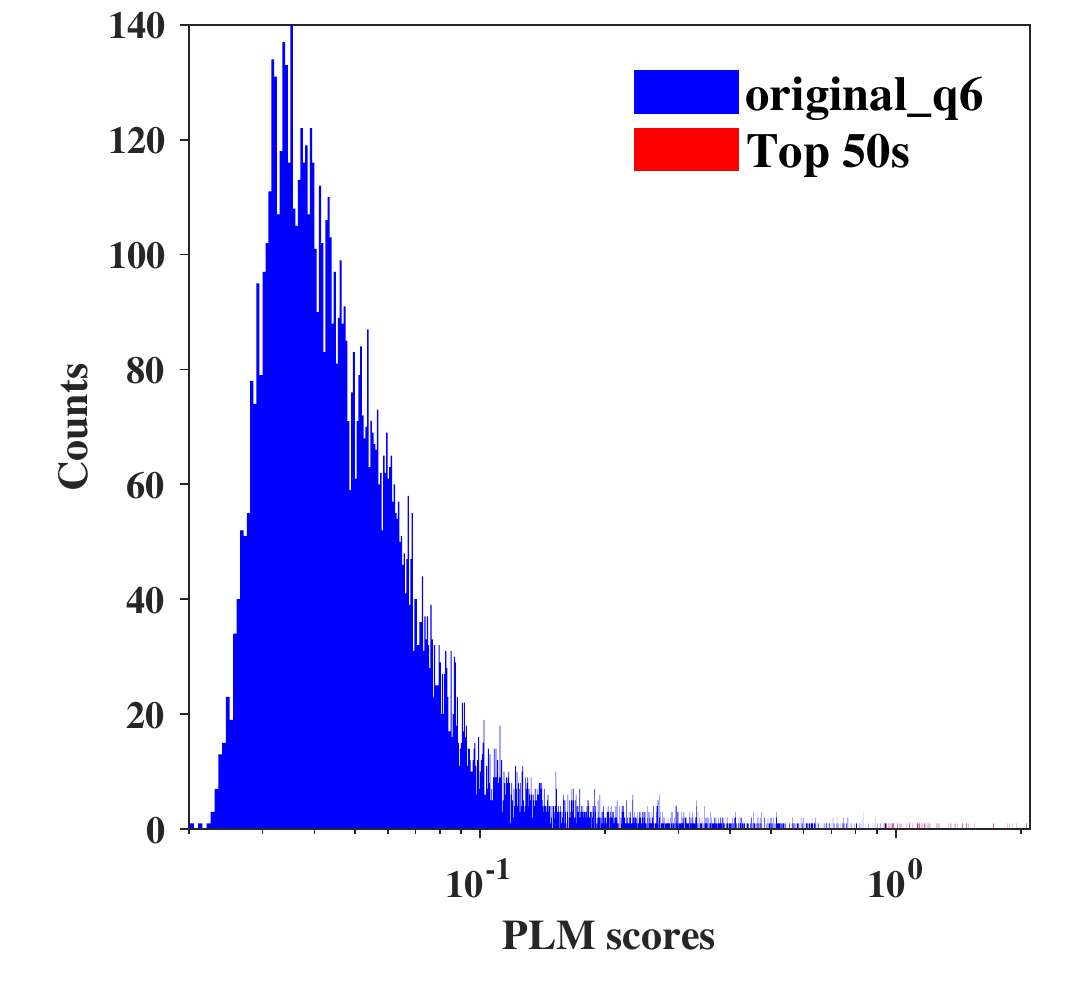}
\put(-162,161){(a)}
{\color{red} \put(-12.1,30){\tiny{$\downarrow$}}} \put(-12.5,33.4){{\tiny{1}}}
{\color{red} \put(-18.5,30){\tiny{$\searrow$}}} \put(-19.5,33.4){{\tiny{2}}}
{\color{red} \put(-24.5,30){\tiny{$\downarrow$}}} \put(-24.8,33.4){{\tiny{9}}}
{\color{red} \put(-31,30){\tiny{$\downarrow$}}} \put(-33,33.4){{\tiny{13}}}
{\color{red} \put(-38,30){\tiny{$\searrow$}}} \put(-42,33.4){{\tiny{22}}}
\put(-42,39){{\tiny{29}}}\\
\includegraphics[width=0.4\textwidth]{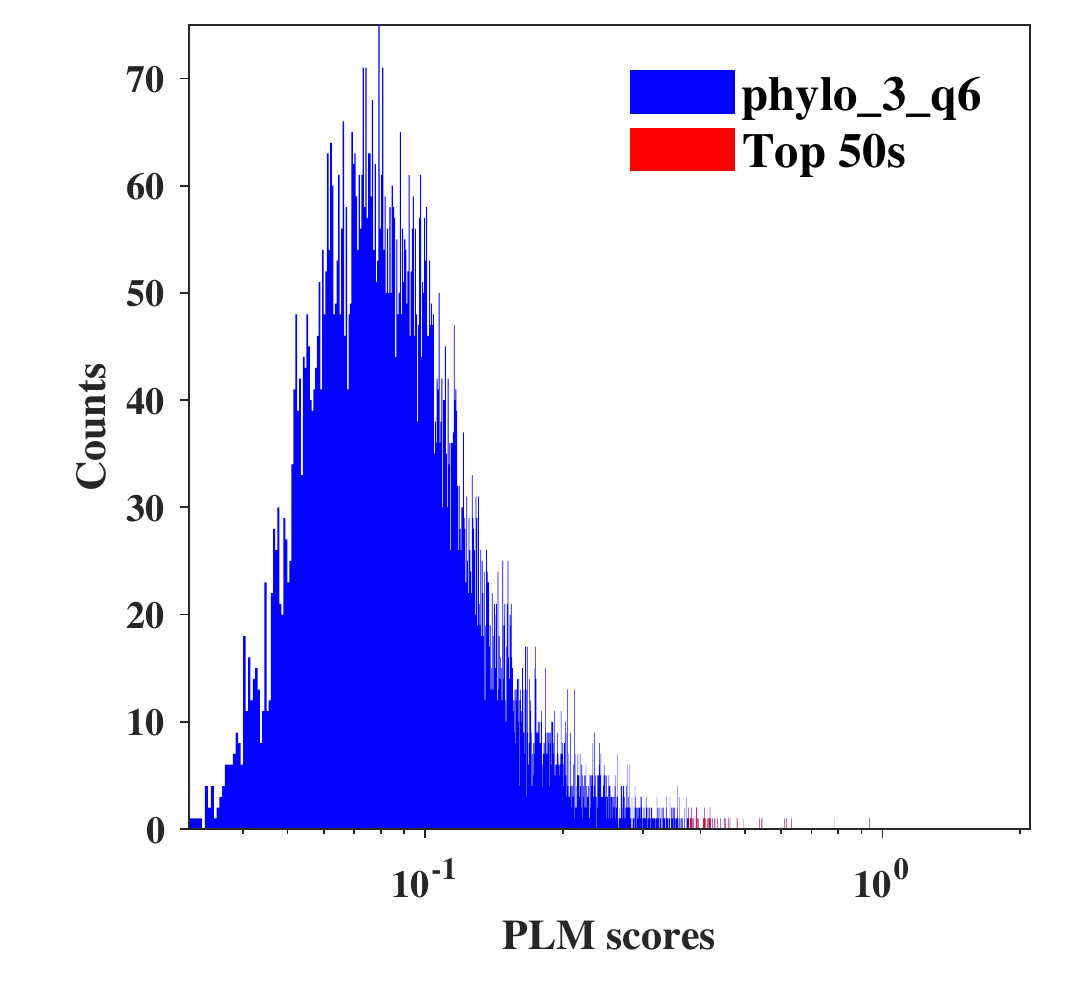}
\put(-162,161){(b)}
{\color{red} \put(-40.8,31){\tiny{$\downarrow$}} }\\
\includegraphics[width=0.4\textwidth]{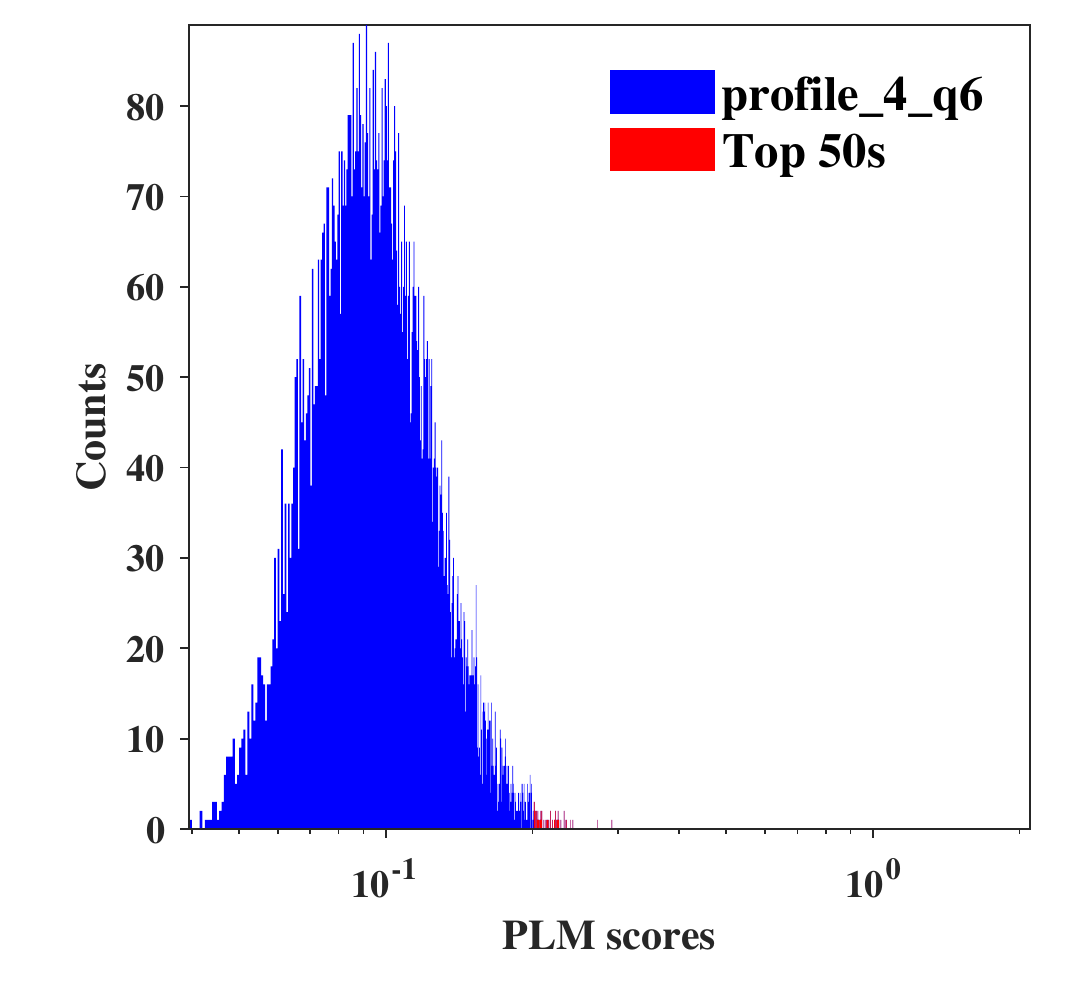}
\put(-162,161){(c)}
{\color{red} \put(-87.3,31){\tiny{$\downarrow$}}}
\caption{Histograms of PLM scores for (a) 2020-05-02 data-set, (b) a phylogeny randomized sample and (c) a profile randomized sample. The blue bars for all scores while the red ones for top-50 largest scores.
Red arrows in (a) indicate links listed in Table~\ref{tab:tab3}, with the 29th link almost overlapping with the 22nd. The largest PLM score is pointed to by red arrows for random samples in (b) and (c). None of them is located inside a coding region, and none of them appear in  Table~\ref{tabs20200502} and Table \ref{tab:tab3} in the main context.
} \label{fig:hists_20200502}
\end{figure}

Table~\ref{tabs20200502} presents the top-50 significant links for the 20200502 data set, all of them appear in Table \ref{tab:tab1} in the main context for the 20200808 data set. Table ~\ref{tab:20200505_tab4} shows the top 10 significant links provided by the correlation analysis for the 20200502 data, which could be compared with Table \ref{tab:tab4} in the main body of the paper. The Fig.~\ref{fig:hists_20200502} shows the histogram of PLM scores for the original MSAs (Fig.~\ref{fig:hists_20200502}(a)) as well as that for phylogeny (Fig.~\ref{fig:hists_20200502}(b)) and profile (Fig.~\ref{fig:hists_20200502}(c)) randomization respectively. This plot can be compared with the Fig. 2 in the main context.

\begin{table*}[htpb]
\centering
\caption{Top-50 significant epistatic links between loci for the 2020-05-02 data-set. To be compared to
Table \ref{tab:tab1} in main body of paper.}
\begin{ruledtabular}
\begin{tabular}{llllll}
\textrm{Rank\footnote{Indices of significant links in the top 50s with both terminals located inside a coding region. Analogous tables for 2020-04-01 and 2020-04-08 data-sets are available on~\cite{Zeng-github}}} & \textrm{Locus 1 -\footnote{Information on the starting locus: index in the reference sequence, the protein it belongs to.}} & \textrm{mutation-\footnote{Information on mutations of the starting locus: the first and second prevalent  nucleotide at this locus, mutation type: synonymous(syn.), non-synonymous(non.).}} & Locus 2 - & mutation- & PLM   \\
~   & protein & type & protein & type & score  \\
\hline
1&~~8782-nsp4 & C$|$T-syn. & 28144-ORF8 & T$|$C-non. & 2.0649 \\
2&~~1059-nsp2 & C$|$T-non.  & 25563-ORF3a & G$|$T-non. & 1.9480  \\
3&~~28882-N & G$|$A-syn. & 28883-N & G$|$C-non. & 1.9116 \\
4&~~28881-N & G$|$A-non. & 28882-N & G$|$A-syn. & 1.8774 \\
5&~~28881-N & G$|$A-non. & 28883-N & G$|$C-non. & 1.8594 \\
9&~~17747-nsp13 & C$|$T-non. & 17858-nsp13 & A$|$G-non. & 1.4798 \\
13&~~11083-nsp6 & G$|$T-non. & 14805-nsp12 & C$|$T-syn. & 1.3876 \\
16&~~3037-nsp3 & T$|$C-syn. & 23403-S & G$|$A-non. & 1.3374 \\
17&~~3037-nsp3 & T$|$C-syn. & 14408-nsp12 & T$|$C-non. & 1.2766 \\
20&~~14408-nsp12 & T$|$C-non. & 23403-S & G$|$A-non. & 1.2101 \\
22&~~17858-nsp13 & T$|$C-non. & 18060-nsp14 & C$|$T-syn. & 1.1973 \\
29&~~17747-nsp13 & C$|$T-non. & 18060-nsp14 & C$|$T-syn. & 1.1392 \\
\end{tabular}\\
\end{ruledtabular}
\label{tabs20200502}
\end{table*}

\begin{table}[htpb]
\caption{Top-10 significant links found by correlation analysis in the coding region for the data set till 2020-05-02. To be compared to
Table 4 in main body of paper.}
\centering
\begin{ruledtabular}
\begin{tabular}{lllc}
\textrm{Rank \footnote{Rank for top-10 coding links highlighted by correlation analysis. Links located within coding regions and the corresponding nucleotide mutation excluding gaps or `N's are considered.}} & Locus 1 - & Locus 2 - & Frobenius \\
~ &protein & protein  & Score\\
\hline
1 & 14408-nsp12 &	23403-S &	0.4726\\
2 & 3037-nsp3   &   23403-S &	0.4711\\
3 & 3037-nsp3   &   14408-nsp12  &	0.4706\\
492 & 1059-nsp2  & 25563-ORF3a & 0.2894 \\
570 & 8782-nsp4 & 28144-ORF8 & 0.2803 \\
776 & 28882-N & 28883-N & 0.255\\
779 & 28881-N & 28882-N & 0.2543\\
782 & 28881-N & 28883-N & 0.2542\\
1309 & 14408-nsp12 & 25563-ORF3a & 0.2086\\
1044 & 23403-S & 25563-ORF3a & 0.2081\\
\end{tabular}\label{tab:20200505_tab4}
\end{ruledtabular}
\end{table}

\bibliography{Covid19}

\end{document}